\documentclass[twocolumn, twocolappendix]{aastex631}

\usepackage{amsmath}
\usepackage{graphicx}
\usepackage{multirow}

\newcommand{\td}[2]{\frac{d #1}{d #2}}

\newcommand{\pd}[2]{\frac{\partial#1}{\partial#2}}

\graphicspath{{./}{figures/}}

\begin{document}

\title{Shock and SEP Modeling Study for the 5 September 2022 SEP Event}



\author[0000-0001-6589-4509]{A. Kouloumvakos}
\affiliation{The Johns Hopkins University Applied Physics Laboratory, \\
11101 Johns Hopkins Road, Laurel, MD 20723, USA.}
\email{Athanasios.Kouloumvakos@jhuapl.edu}

\author[0000-0001-6344-6956]{N. Wijsen}
\affiliation{Centre for mathematical Plasma Astrophysics, KU Leuven Campus Kulak, 8500 Kortrijk, Belgium}

\author[0000-0001-6344-6956]{I. C. Jebaraj}
\affiliation{Department of Physics and Astronomy, University of Turku, FI-20500 Turku, Finland}

\author[0000-0001-9325-6758]{A.~Afanasiev}
\affiliation{Department of Physics and Astronomy, University of Turku, FI-20500 Turku, Finland}

\author[0000-0002-3176-8704]{D. Lario}
\affiliation{Heliophysics Science Division, NASA Goddard Space Flight Center, Greenbelt, MD 20771, USA.}

\author[0000-0002-0978-8127]{C. M. S. Cohen}
\affiliation{California Institute of Technology, 1260 E California Blvd, Pasadena, CA 91125, USA.}

\author[0000-0002-1859-456X]{P. Riley}
\affiliation{Predictive Science Inc. (PSI), 9990 Mesa Rim Road, Suite 170, San Diego, CA 92121, USA.}

\author[0000-0003-1960-2119]{D. G. Mitchell}
\affiliation{The Johns Hopkins University Applied Physics Laboratory, \\
11101 Johns Hopkins Road, Laurel, MD 20723, USA.}

\author[0000-0002-9829-3811]{Z. Ding}
\affiliation{Institute of Experimental and Applied Physics, Kiel University, 24118 Kiel, Germany}

\author[0000-0002-8164-5948]{A. Vourlidas}
\affiliation{The Johns Hopkins University Applied Physics Laboratory, \\
11101 Johns Hopkins Road, Laurel, MD 20723, USA.}

\author[0000-0002-0850-4233]{J. Giacalone}
\affiliation{Lunar and Planetary Laboratory, University of Arizona, Tucson, AZ, USA.}

\author[0000-0003-2865-1772]{X. Chen}
\affiliation{University of Michigan, 500 S. State Street, Ann Arbor, MI 48109, USA.}

\author[0000-0002-5674-4936]{M. E. Hill}
\affiliation{The Johns Hopkins University Applied Physics Laboratory, \\
11101 Johns Hopkins Road, Laurel, MD 20723, USA.}


\begin{abstract}
On September 5, 2022, during Parker Solar Probe's (PSP) 13th encounter, a fast shock wave and a related solar energetic particle (SEP) event were observed as the spacecraft approached the perihelion of its orbit. Observations from the Integrated Science Investigation of the Sun (IS$\sun$IS) instrument suite show that SEPs arrived at the spacecraft with a significant delay from the onset of the parent solar eruption and that the first arriving SEPs exhibited an Inverse Velocity Dispersion (IVD) for energetic protons above $\sim$1~MeV. Utilizing data from multiple spacecraft we investigate the eruption dynamics and shock wave propagation. Our analysis includes 3D shock modeling and SEP transport simulations to examine the origins of this SEP event and explore the causes of the delayed SEP onset and the observed IVD. The data-driven SEP simulation reproduces the SEP event onset observed at PSP, its evolving energy spectrum and the IVD. 
This IVD is attributed to a relatively slow, ongoing particle acceleration process occurring at the flank of the expanding shock wave intercepted by PSP. 
This has significant implications for the role of shocks in the release of SEPs at widespread events and for methods used to infer the SEP release times. 
Furthermore, the match between the simulation and observations worsens when cross-field diffusion is considered, indicating that SEP diffusion had a minor effect on this event. These findings underscore the complexity of SEP events and emphasize the need for advanced modelling approaches to better understand the role of shock waves and other physical processes in SEP acceleration and release.
\end{abstract}


\section{Introduction} \label{sec:intro}

During large solar eruptions such as flares and coronal mass ejections (CMEs), charged particles can be accelerated to high energies (above a few GeV close to the Sun) \citep[e.g.][]{Reames2009}. The underlying acceleration process is yet to be fully understood. These Solar Energetic Particles \citep[SEPs; see reviews by][]{Reames1999, Desai2016} can fill up the inner heliosphere within minutes. Some of these events can be extremely intense and cause various impacts on satellites \citep{Buzulukova2022, Pirjola2005}. They can also pose an extreme threat to manned space missions \citep{Patel2020}. SEP research is, therefore, essential in the rapidly expanding field of space human exploration. Understanding the mechanisms that accelerate and transport energetic particles in the heliosphere has been a top scientific goal for over six decades of solar physics research \citep[see for example][and references therein]{Reames2021, Reames2022} and it is also an important element of space weather forecasting \citep{Whitman2023}.

Multi-spacecraft observations of SEPs at various heliospheric locations have greatly enhanced our understanding of the origin and transport of these particles \citep{Rouillard2012, Lario2014}. The most intense SEP events are accompanied by flares and CMEs that drive fast, wide, and strong shock waves \citep{Cliver2004, Reames2009, Papaioannou2016}. Shock waves can accelerate charged particles through several different mechanisms \citep[e.g.,][]{Vainio2018}. A widely accepted mechanism for proton acceleration during large SEP events is diffusive shock acceleration \citep[DSA;][]{Krymskii77, Axford1977, Bell1978, Blandford1978} where charged particles gain energy by crossing the shock front multiple times \citep[e.g. first-order Fermi acceleration mechanism][]{Fermi1954}. During these intense SEP events, the associated CME-driven shocks reach a high shock strength \citep{Rouillard2016, Kouloumvakos2019}. The shock strength and geometry can vary significantly along the shock front surface \citep[e.g.][]{Pomoell2011, Jin2018, Jebaraj2021} and can greatly influence the efficiency of the particle acceleration mechanisms at shocks \citep[e.g.][]{Giacalone2012, Caprioli2014, Giacalone2017, Jebaraj2023, Afanasiev2018, Chen2022, wijsen2023ApJ, Jebaraj23L}. Moreover, the interaction of the shock with the denser coronal streamers can lead to efficient acceleration of particles at the early stages of the solar events, when the associated shocks are still low in the corona \citep{Kong2017, Kong2019, Kouloumvakos2020, Frassati2022, Liu2023}. Lastly, modeling of diffusive shock acceleration has shown that strong CME-driven shocks can accelerate protons from a few hundred keV to several GeV in a matter of minutes under certain conditions \citep[e.g.][]{Ng2008, Afanasiev2018}. 

The spread of SEPs across radial and angular distances can be due to a combination of different processes \citep{Lario2017, Rodriguez2021, Kouloumvakos2022}, including the extent of the shock wave \citep{Cane1988, Rouillard2012, Lario2014, GomezHerrero2015, Zhu2018, Kouloumvakos2015}, SEP transport processes such as perpendicular (cross-field) diffusion \citep[e.g.][]{Dalla2003, Droge2010,shalchi2020, strauss2020, Wang2023}, and in some rare occasions distinct SEP injections \citep{Dresing2023}. In a recent study, \cite{Strauss2023} showed that a combination of a relatively broad particle source such as a wide CME/shock together with SEP perpendicular diffusion transport can successfully describe the SEP onset times observed by multiple spacecraft in many events. Other studies have shown that a rapidly expanding shock wave higher in the low corona is able to explain the observed release of SEPs to distant observers \citep[e.g.][]{Zhu2018}, for some widespread events. \cite{Kouloumvakos2023} suggested that the varying shock properties at the magnetic field lines connected to an observer could be an important factor in determining the observed timings with respect to the parent solar eruption \citep[see also][]{Kihara2023}.

Despite the wealth of multipoint observations in the inner heliosphere and the advancements in numerical models, the origin of high-energy particles and the relative contributions of various acceleration processes (e.g. shocks-related such as DSA or flare-related such as magnetic reconnection) remain unclear. Moreover, it is uncertain under which conditions large-scale shock waves can efficiently inject SEPs to distant magnetically connected spacecraft and when a combination of multiple processes may become necessary for an efficient longitudinal dispersal of SEPs. Observations from the Parker Solar Probe \citep[PSP;][]{Fox2016} and Solar Orbiter \citep[][]{Muller2020}, close to the Sun, offer promising insights into these unresolved questions, particularly regarding the mechanisms primarily responsible for accelerating and transporting energetic particles in the heliosphere and their relative contributions.

On September 5, 2022, a fast and large CME erupted from the far side of the Sun as seen from Earth, which was associated with a broad shock wave and an M9 class flare \citep[see][]{Long2023}. The CME was one of the fastest CMEs of Solar Cycle 25 at that time with a speed of $\sim$2039~km/s \citep[see][]{Paouris2023} and \cite{Patel2023} showed that this CME is among the fastest 0.15\% of all CMEs listed in the CDAW CME catalog\footnote{\url{https://cdaw.gsfc.nasa.gov/CME_list/}}. The accompanying SEP event was one of the most intense observed throughout the Solar Orbiter mission at that time and it was characterized by a strong energetic storm particle (ESP) event (e.g. a rapid increase in the particle intensities associated with the CME-driven shock passage), recently analyzed by \citet{Ding2024}. For PSP, it was the first time that a very fast CME was observed at such a close radial distance from the Sun. An analysis of the shock wave observed in situ at PSP and Solar Orbiter is presented in \citet{Trotta2024}. A detailed overview of the PSP SEP measurements from the Integrated Science Investigation of the Sun \citep[IS$\Sun$IS;][]{McComas2016} is presented in \cite{Cohen2024} that shows many interesting aspects of this event, such as an inverse velocity dispersion in the arrival of protons above $\sim$1 MeV. More recently, \cite{Jebaraj24b} demonstrated that the shock was also an emitter of synchrotron radiation, an exceptionally rare energetic phenomena in heliospheric shocks. 

In this study, we present and analyze in Section 2 SEP measurements from PSP and remote sensing observations during the event. In Section 3, we perform a detailed 3D modelling of the shock wave in the corona and combine the modeling results with an SEP model to simulate the SEP production at the shock. Then we compare the simulation results with SEP observations made by PSP very close to the Sun and we examine the role of the shock wave to the acceleration and release of the SEPs observed by the spacecraft. In Section 4, we discuss the results of (1) the cause of the delayed SEP onset and (2) the origin of the inverse velocity dispersion that was observed by PSP and summarize our findings.

\begin{figure}
  \centering
  \hspace{-1em}\includegraphics[width=0.47\textwidth]{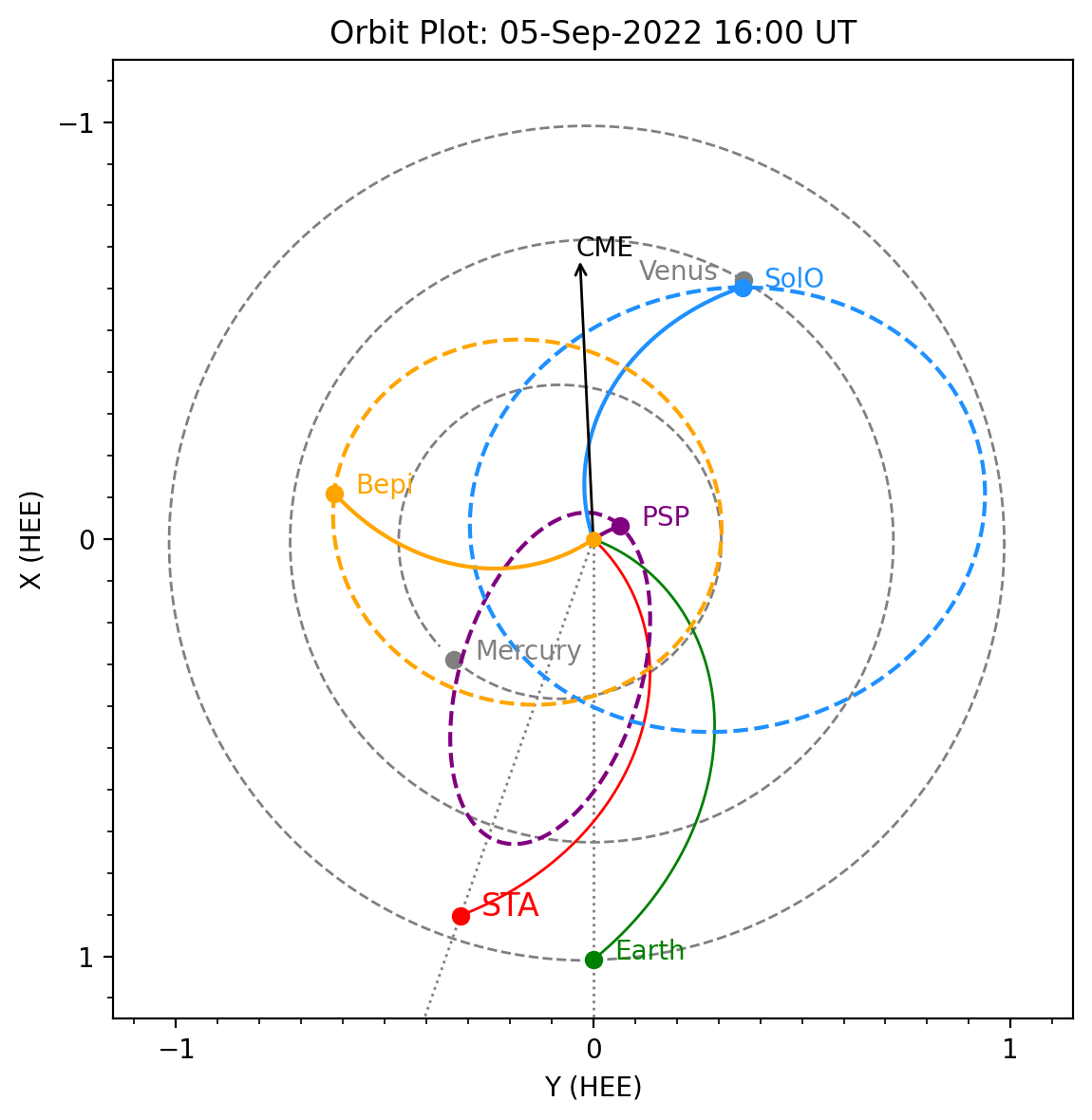}
\caption{A view of the ecliptic that shows the positions of the planets and spacecraft in the inner heliosphere on September 5, 2022, at 16:00 UT. The orbits of the spacecraft(planets) are shown with the colored(grey) dashed lines. The Parker spiral connecting the different observers to the solar surface are shown with the solid colored lines. The arrow marks the direction of propagation of the CME. The orange circle at the center indicates the Sun (not in true scale).}
     \label{fig:orbits_view}
\end{figure}

\section{Observations} \label{sec:observ}

The solar event associated with the origin of the SEP event on 2022 September 5 was one of the first extreme solar events of Solar Cycle 25, i.e.  a fast CME driving a strong shock resulting in an intense widespread SEP event. It was observed by a large fleet of spacecraft, widely distributed in the inner heliosphere. In Figure~\ref{fig:orbits_view}, we show a view of the ecliptic with the positions of the planets and spacecraft near the onset of the eruption and the direction of the CME. At the time of the parent solar eruption (estimated to be at $\sim$16:15 UT), PSP was at a heliocentric radial distance of 15.4~R$_\sun$, almost at the perihelion of its orbit \#13, and Solar Orbiter was at 151.1~R$_\sun$. 

In our study, we use remote sensing EUV, coronagraphic, and heliospheric observations of the solar corona obtained from multiple spacecraft. These include images from the Extreme Ultraviolet Imager \citep[EUVI;][]{Wuelser2004} and the COR2 white-light coronagraph both part of the SECCHI instrument suite \citep{Howard2008} onboard STEREO-A which at the time was trailing Earth. Additionally, images from the Large Angle and Spectrometric COronagraph \citep[LASCO;][]{Brueckner1995} C2 on board Solar and Heliospheric Observatory \citep[SOHO][]{Domingo1995} which orbits around the first Lagrange point (L1). We also use EUV images from the Atmospheric Imaging Array \citep[AIA;][]{Lemen2012} onboard the Solar Dynamics Observatory \citep[SDO;][]{Pesnell2012} at a geosynchronous transfer orbit, and images from the Extreme Ultraviolet Imager \citep[EUI;][]{Rochus2020} instrument onboard Solar Orbiter. We also utilized SEP measurements from the IS$\Sun$IS energetic particle suite on PSP and the Energetic Particle Detector \citep[EPD;][]{Pacheco2020} instrument suite on Solar Orbiter. IS$\Sun$IS suite comprises two Energetic Particle Instruments (EPI) that measure low \citep[EPI-Lo;][]{Hill2017} and high (EPI-Hi) energy particles. EPI-Lo provides measurements of energetic ions from $\sim$20 keV/nucleon to 10~MeV total energy, and EPI-Hi measures ions from $\sim$1 to 90 MeV/nucleon. EPD instrument suite on Solar Orbiter comprises four telescopes \cite[see also][]{Wimmer2021}. We use observations of energetic ions from the Electron Proton Telescope (EPT) in the energy range 
from 20~keV to 15~MeV and the High Energy Telescope (HET) from 7~MeV to 107~MeV.

\subsection{Remote-sensing observations}

The eruption on September 5, 2022, took place on the far side of the Sun from Earth's viewpoint. Therefore, near-Earth observations in EUV could not offer any view of the erupting active region (AR) or the CME lift-off. The AR 13088, where the solar event took place, was only visible from the Solar Orbiter's viewpoint. The Full Sun Imager (FSI) of EUI observed the eruption with a cadence of about ten minutes per image \citep[see][]{Long2023}. On the other hand, coronagraphic observations from STEREO-A give a good view of the CME evolution in the corona when it was already over the limb, as seen by this spacecraft. At the time of the event, STEREO-A COR2 was in a high-cadence campaign and provided coronagraphic images at a 5-minute cadence. Wide-Field Imager for Parker Solar Probe \citep[WISPR;][]{Vourlidas2016} observed in great detail the CME evolution and its internal structure, which was more complex than what was observed from the 1~au view.

\begin{figure*}
  \centering
  \includegraphics[width=0.9\textwidth]{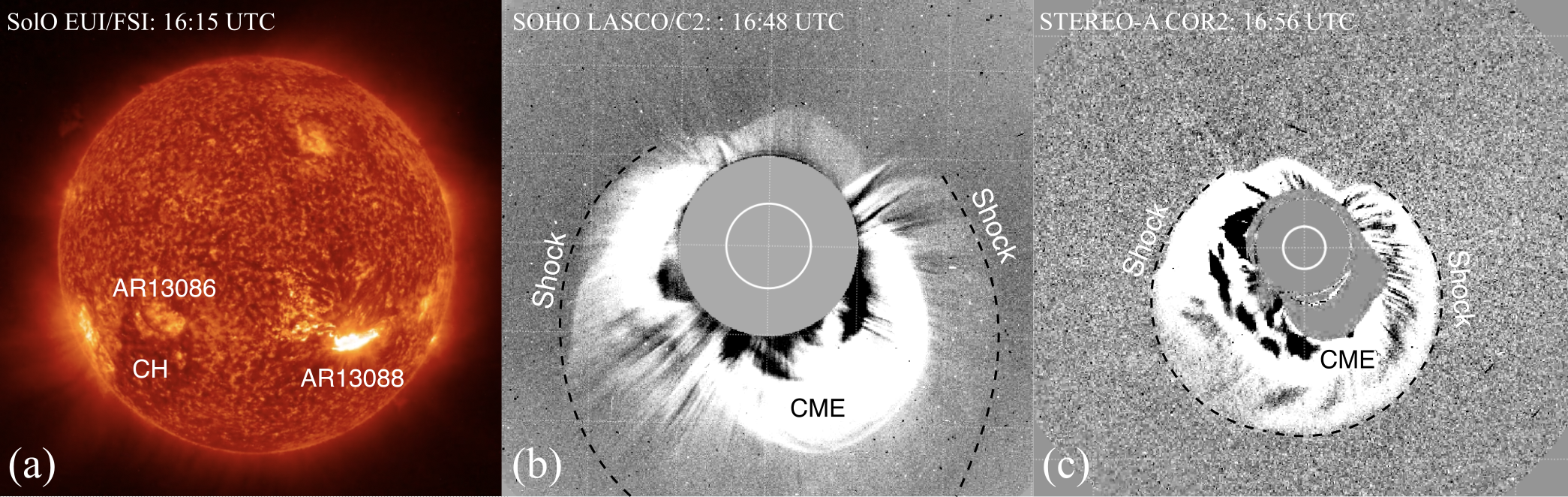}
\caption{Remote sensing observations during the eruption. Panel (a) shows a plain image from EUI/FSI at 304~\AA. Panel (b) and (c) show running-difference white-light coronagraphic images of the CME and the shock from LASCO/C2 and STEREO-A/COR2, respectively. We label on panel (a) the location of AR 13088 and a coronal hole located at the south-eastern hemisphere, and on panels (b) and (c) the CME and the white-light shock.}
     \label{fig:remote_senc}
\end{figure*}

Figure~\ref{fig:remote_senc} shows EUV and white light observations. Panel~(a) shows an image at 304~\AA~from EUI/FSI at 16:15~UTC. The bright flare ribbons are visible at the bottom right of the EUI image, where the AR 13088 was located. The solar event in EUV is complex, with multiple brightenings observed at the flaring region before and after the CME liftoff. This is in accordance with the hard X-ray observations from Solar Orbiter STIX that show multiple flare episodes (Vievering et al. 2024, under review).

The CME liftoff is accompanied by an EUV wave in the low corona (Long et al. 2023), which is considered to be a CME-driven fast-magnetosonic wave \citep[e.g.][]{Warmuth2015, Patsourakos2012, Long2017}. The wave is barely visible in the Solar Orbiter images at 174~\AA, since this EUV bandpass is not sensitive to hot plasma emission that is present in the pressure wave and the observing cadence is low. As the disturbance propagates in the low corona, it causes the displacement of various coronal structures, leaving faint but visible traces. The EUV wave is also visible off-limb in STEREO-A EUVI images for a few frames, as a bright front that propagates off-disk at a position angle of 180$^\circ$ \citep[see Figure~2 in][]{Paouris2023}.

This event is also accompanied by a broad and fast CME and a white-light shock wave \citep[see examples in][]{Ontiveros2009}. Panels~(b) and~(c) of Figure~\ref{fig:remote_senc}, show the well-developed CME and shock wave in coronagraphic images from LASCO/C2 and STEREO-A/COR2, at 16:48~UTC and 16:56~UTC, respectively. The shock wave in the white-light coronagraphic images can be seen as a bright front around the CME. We outline the shock front at the two running difference coronagraphic images of Figure~\ref{fig:remote_senc}. The CME, on the other hand, is the bright core structure inside the shock envelope. That the shock wave was very broad can be seen from the deflected streamers that are located at the image plane of the two coronagraphic images, almost 90$^\circ$ away from the eruption site. This extended propagation of the shock on the opposite side of the eruption is observed in other extreme events. \cite{Kwon2017} have shown that under some conditions, strong shock waves can even encompass the whole Sun and become global shocks. The CME and shock wave were particularly fast. \cite{Paouris2023} showed that the southern portions of the shock and CME displayed the highest plane-of-sky speeds. More specifically, based on \cite{Paouris2023} kinematic analysis from COR2A/STEREO images, the shock's maximum speed at the plane-of-sky was found to be $\sim$2260~km/s.

\subsection{Solar energetic particle observations}

\begin{figure}
  \hspace{-0.10in}
  \includegraphics[width=0.5\textwidth]{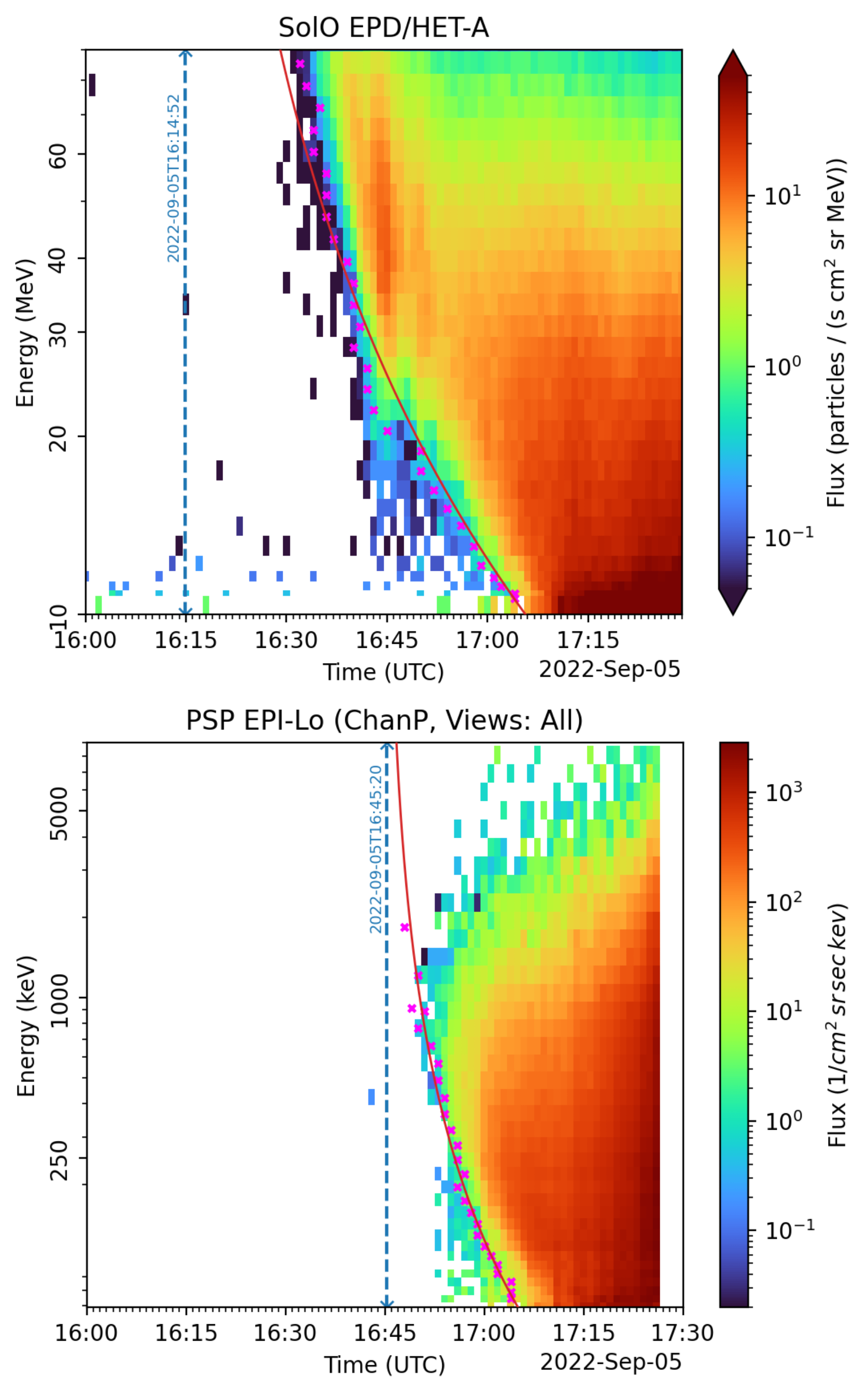}
\caption{Observations of solar energetic protons from Solar Orbiter (top panel) and PSP (bottom panel). Top panel: shows the energy spectrogram of proton fluxes from HET-Sun telescope on Solar Orbiter at an energy range from $\sim$10~MeV to $\sim$90~MeV. Bottom panel: shows the energy spectrogram of proton fluxes from the Epi-Lo instrument on PSP. This spectrogram incorporates proton data from all viewing directions of Epi-Lo. At each panel, the red line is the velocity dispersion curve fitted to the SEP onset times depicted with the magenta crosses. The vertical dashed lines at each panel denote the SEP release times at each spacecraft from the VDA. The release times have been time shifted from the Sun to the spacecraft taking into account the light travel time to each observer.}
     \label{fig:SEP_Observations}
\end{figure} 


In association with this big solar eruption, multiple spacecraft observed an SEP event. Figure~\ref{fig:SEP_Observations} shows SEP observations from Solar Orbiter EPD (top panel; HET) and PSP IS$\sun$IS/ (bottom panel; EPI-Lo). At Solar Orbiter, the SEP event shows a prompt onset and a clear velocity dispersion in proton arrival times. This can be seen from the top panel of Figure~\ref{fig:SEP_Observations} that shows the energy spectrogram of proton fluxes from the sunward-looking HET telescope (HET-Sun) on Solar Orbiter. The event was both intense and long-lasting, with high-energy protons intensities (above 80~MeV) remaining elevated above background levels for more than two days. One reason for the high intensity of the SEP event at Solar Orbiter is the spacecraft's close magnetic connection to both the source active region (within 10$^\circ$ longitude) and the apex of the propagating CME-driven shock, as discussed in Section 4. On the other hand, PSP observed the SEP event from a closer solar distance. PSP magnetic connection was estimated to be close to AR 13086, which had emerged within a coronal hole a few days before the event (see Figure~\ref{fig:remote_senc}). EPI-Lo observations above ~1 MeV show an inverse velocity dispersion pattern, where higher-energy protons arrived later than those of lower energy. This is evident in the bottom panel of Figure~\ref{fig:SEP_Observations} that shows the energy spectrogram of proton fluxes from EPI-Lo instrument on PSP. A detailed analysis of this remarkable PSP observation is elaborated further in the analysis of IS$\sun$IS EPI data by \cite{Cohen2024}. This study showed that particles with energy above 1~MeV arrive later than the particles with energy below 1~MeV. \cite{Cohen2024} also showed that there is a sharp decrease in the SEP intensities at all energies after the shock crossing at 17:27~UT and repeated short durations of highly anisotropic sunward flows. From the EPI-Hi LET observations we find that the onset of the event at PSP is at around 16:48~UTC on September 5, 2022, for $\sim$1.8~MeV protons. This onset time is notably delayed compared to Solar Orbiter, where the first protons observed at HET-Sun (with energies of ~85 MeV) arrived at around 16:32~UTC, with energies of $\sim$85~MeV. Given that Solar Orbiter was located farther from the Sun than PSP, this significant time difference suggests that the release of SEPs observed by PSP was substantially delayed.

We estimate the SEP release times at each spacecraft using the velocity dispersion analysis \citep[VDA; e.g.,][]{Vainio2013JSWSC}. This method is based on the assumptions that the first arriving particles to the observer have been released simultaneously and experience no scattering during their interplanetary transport along a common path length \citep[see more details in][]{Laitinen2015}. Then if these conditions are true, the particles’ onset times (t$_0$) at the observer are given by

\begin{equation}
    t_0 = t_{rel} + \frac{L}{\upsilon}
\end{equation}

where $t_{rel}$ is the release time at the Sun, \(L\) is the path length traveled by the particles, and $\upsilon$ is the particles’ speed. The SEPs release time at the Sun and the path length traveled can be calculated by a linear fitting of the observed onset times at different energies. We applied the VDA method to SEP measurements by the HET instrument on Solar Orbiter and EPI-Lo and EPI-Hi on PSP. For Solar Orbiter we used data from the HET-Sun telescope, at an energy range of $\sim$10~MeV to $\sim$90~MeV and for PSP we used data from EPI-Lo using all viewing directions and EPI-Hi from the sunward looking LET telescope (LET-A) at an energy range from $\sim$70~keV to $\sim$9~MeV and $\sim$0.9 to $\sim$2~MeV, respectively. For the VDA at PSP we used only the onset times at an energy lower than the energy where the inverse velocity starts to form.

From the VDA, we find an SEP release time at 16:09$\pm$4min UTC for Solar Orbiter and at 16:45$\pm$1min UTC for PSP. It is important to note that these release times are calculated at the Sun and can be compared with each other. To determine the corresponding release times at the spacecraft (or at 1~au) and compare them with electromagnetic observations at each location, we need to account for the light travel time from the release site and add it to the release times. Comparing the release times estimated for the two spacecraft, we find that the SEPs were released with a time difference of about 35 minutes confirming that there is a significant delay. For the particle travel path length, we find L$\sim$0.98$\pm$0.08~au for Solar Orbiter and L$\sim$6.3$\pm$0.6~R$_\sun$ for PSP. Focusing on PSP, the path length is much shorter than the nominal Parker spiral length. This is an important result which suggests that the SEPs may not have been released near the solar surface but higher in the corona at a distance of approximately 9 R$_\sun$. Therefore, the expanding shock and the evolution of the shock parameters at the magnetic field lines connected to the spacecraft could have played an important role in the observed properties of the SEP release at PSP \citep[see e.g.][]{Kouloumvakos2023} and we will explore this further in our study.

\section{Modeling} \label{sec:Modeling}

In this study, we perform a detailed 3D modelling of the shock wave in the corona using \textit{PyThea} \citep[][]{Kouloumvakos2022_Pythea}, an open-source software package in Python available online on GitHub and Zenodo\footnote{\url{https://doi.org/10.5281/zenodo.5713659}} that provides tools to reconstruct the 3D structure of CMEs and shock waves, and methods presented in \cite{Kouloumvakos2019} to model the shock parameters in 3D. Then we combine the shock model with an SEP model, named Particle Radiation Asset Directed at Interplanetary Space Exploration \citep[PARADISE;][]{Wijsen2019, Wijsen2020}, which we couple to a coronal model (Magnetohydrodynamic Algorithm outside a Sphere,\citep[MAS;][]{Mikic1999, Riley2001}) and heliospheric model (European heliospheric forecasting information asset \citep[EUHFORIA][]{Pomoell2018, Poedts2020}). From all the above we construct a data-driven SEP model in the solar corona.

These latter two magnetohydrodynamic (MHD) models provide a description of the background plasma, serving as the medium through which the shock and the energetic particles are propagated. This particular combination of models is among the first attempts to model particles accelerated early in a solar event when the shock was still located in the low corona and presumably capable of efficiently accelerating particles \citep[e.g.][]{Gopalswamy2005, Reames2009}. We use a similar approach as \citet{Zhang2023}, who likewise created a data-driven physics-based particle transport model to calculate the SEP acceleration and transport from a CME-driven shock through the solar corona and interplanetary space \citep[see also][]{Li2021}. 

\subsection{Shock reconstruction and kinematics}

We estimate the three-dimensional structure and kinematic properties of the shock wave using an ellipsoid model \citep{Kwon2014}. This geometrical model has been widely used to model the global large-scale structure of CME-driven shocks in the solar corona. The ellipsoid model is defined by three positional parameters that adjust the longitude, latitude, and height of the center and three geometrical parameters that adjust the length of the three semi-axes. To fit the geometrical model near-simultaneous multi-viewpoint observations of the shock are usually utilized. In this study, we used remote sensing coronagraph images from two viewpoints, one provided by STEREO-A and the other from SOHO and SDO. For the fitting process, we used running-difference images. We iteratively adjust the free parameters of the ellipsoid until there is a good visual fit between the model and the observations.

\begin{figure*}
  \centering
  \includegraphics[width=0.98\textwidth]{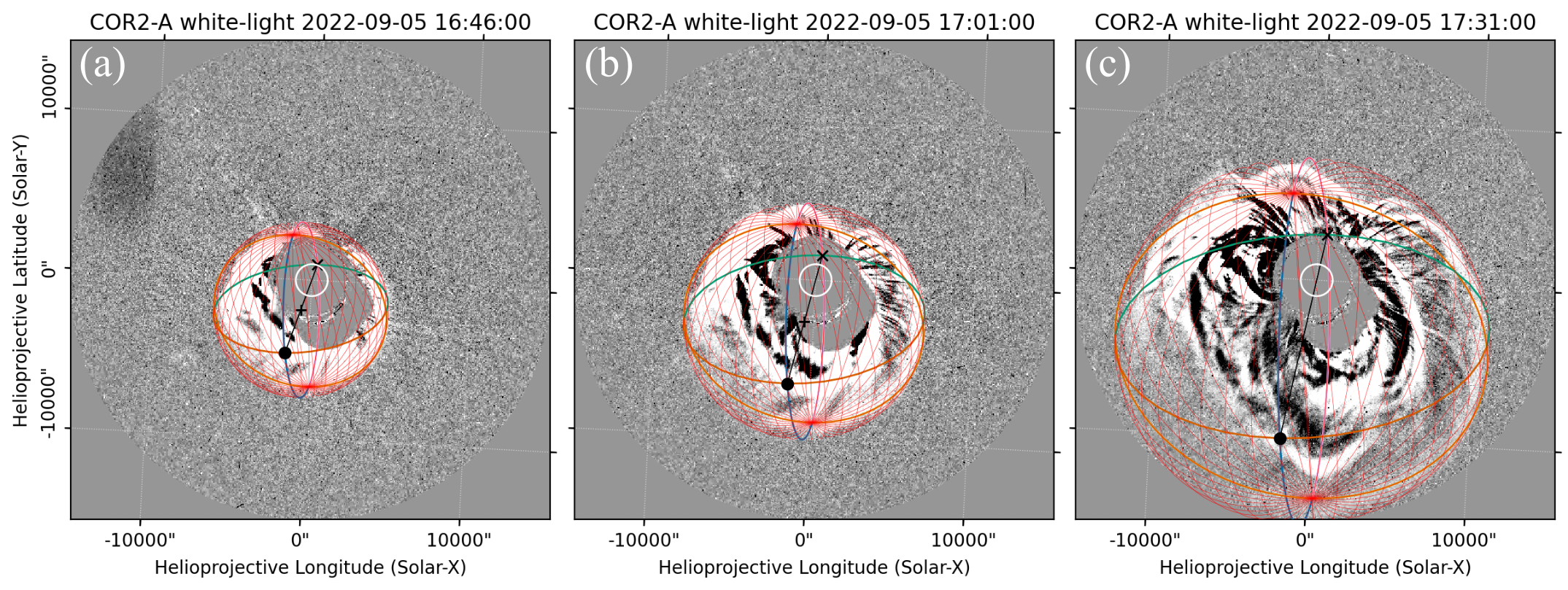}
\caption{ STEREO-A COR2 running-differences images. The reconstructed shock surface is shown at each frame with the red wiremesh. The shock apex and center are marked with the black circle and cross, respectively.}
     \label{fig:shock_reconstruct}
\end{figure*}

The reconstruction of this event was complicated because STEREO-A and Earth's viewpoint were close to each other ($<$20 degrees apart in longitude). We start the shock reconstruction utilizing EUV images from STEREO-A EUVI at 16:10~UTC, when the shock apex was at a radial distance of $\sim$1.9~$R_\sun$. Subsequently, we continued the fitting process primarily using STEREO-A COR2 images from 16:26 to 17:56~UTC, covering a range of shock apex positions from 4.6 to 23.7~$R_\sun$. Fortunately, COR2 had a high-cadence campaign during this period, providing images at five-minute intervals, enabling more precise determination of the shock's kinematics. Throughout the fitting process, we complemented COR2 observations with LASCO/C2 images.
Figure~\ref{fig:shock_reconstruct} shows the wireframe of the ellipsoid fit overlaid in the white-light coronagraphic running-difference images from COR2. The geometric model aligns well with COR2 observations of the white-light shock.

From the shock fitting process, we observe that the positional parameters of the ellipsoid remain nearly constant throughout the entire fitting duration. Specifically, the ellipsoid center is located at a longitude of 170$^\circ$ and a latitude of -41$^\circ$ in the Stonyhurst heliographic coordinate system\footnote{The Stonyhurst heliographic coordinate system has origin at the center of the Sun. The z-axis is aligned with the Sun’s north pole and the x-axis is towards the point of intersection of the solar equator and the solar central meridian as seen from Earth.}. Regarding the geometrical parameters of the shock, we find mean values of approximately $\bar{\kappa}\sim0.65$ for the self-similar constant (defined as the ratio of the apex height to the length of the second semi-axis of the ellipsoid), $\bar{\alpha}\sim1.03$ for the aspect ratio (the ratio of the second to the third semi-axis), and $\bar{\epsilon}\sim-0.27$ for the eccentricity of the ellipsoid \citep[see Eq. 1 in][]{Kouloumvakos2022_Pythea}.

\begin{figure}
  \centering
  \includegraphics[width=0.5\textwidth]{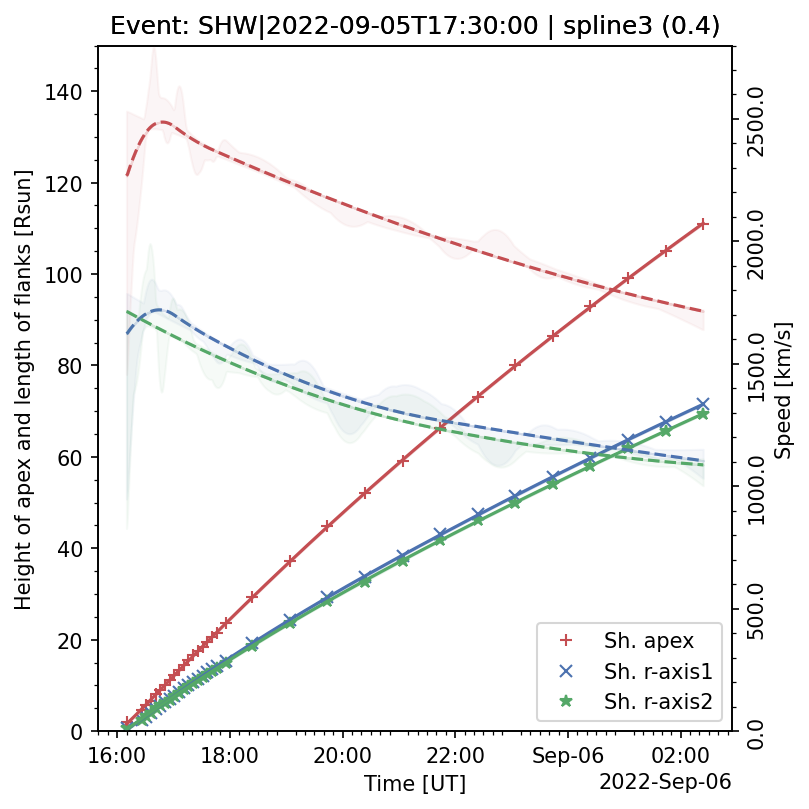}
\caption{Kinematic time profiles of the reconstructed shock wave. The two red curves show the shock kinematics at the apex, the blue and green curves show the kinematics of the flanks. The solid lines show the height(length) and the dashed lines the speed of the apex(flanks). The uncertainties in the shock speed are shown with the color shaded areas.}
     \label{fig:shock_kinematics}
\end{figure}

The results of the 3D shock reconstruction indicate rapid propagation of the shock into the heliosphere. Figure~\ref{fig:shock_kinematics} presents the outcomes of the kinematic analysis. The symbols represent the height of the apex and the length of the two semi-axes, while the colored line passing through the symbols depicts third-order spline fittings at these points. These spline fittings are used to derive the kinematic curves and estimate the propagation and expansion speeds of the shock apex and the two semi-axes.
The shock apex propagates fast in the low corona ($\lesssim$20 R$_{\sun}$). We find that the maximum speed at the apex is $\sim$2480$\pm$150~km/s and at the flanks $\sim$1745$\pm$100~km/s. The shock apex initially accelerates for about 15~min, it reaches the maximum speed at $\sim$16:46~UTC and then decelerates until the end of our fittings.

As mentioned earlier, the direction of the associated CME and the position of STEREO-A and SOHO made it difficult to accurately constrain the shock location near the apex. For this reason, we assessed the accuracy of our fittings through various methods. Initially, we compared the location of the fitted ellipsoid relative to PSP, which observed the eastern flank of the shock in situ very close to the Sun (e.g. at a radial distance of $\sim$15.5~$R_\sun$ at 17:27 UT). Our analysis shows a difference in the arrival time of the shock at the spacecraft of less than two minutes. This suggests that, in this specific direction, our reconstruction represents reasonably well the shock location. However, the modeled speed of the shock flank (\(\sim\)1830 km s\(^{-1}\)) is found to be somewhat larger than the speed estimated by the in situ studies of the event at PSP \citep[\(\sim\)1500 km s\(^{-1}\), ][]{Romeo2023, Trotta2024}. This discrepancy is expected, as local in situ observations are influenced by inhomogeneities in the surrounding medium, which can affect the local characteristics of the shock. Therefore, the shock speed estimated in larger scales from the reconstruction may naturally not match the shock speed measured at smaller scales in situ. Another possibility is that the shock speed from the 3D reconstruction is overestimated so the final accuracy of the kinematics at this direction is about 16\%.


We also compared the location of the fitted ellipsoid with the in situ arrival time of the shock at Solar Orbiter. To achieve this, we extended the shock fitting above 30~$R_\sun$, utilizing observations from the heliospheric imagers HI1 on STEREO-A from 18:23~UTC to 02:23~UTC on September 6, when the shock apex reached $\sim$111~$R_\sun$ and the wavefront became too faint to unambiguously estimate its location. At 02:23~UTC the shock was located at a radial distance of $\sim$95~$R_\sun$ and was propagating with a speed of $\sim$1450~km/s towards Solar Orbiter. Assuming a constant speed until the shock reached Solar Orbiter, we estimated a transit time of $\sim$7.5~hours. After correcting for the light travel time between the remote sensing data at 1~au and in situ measurements at Solar Orbiter, we calculated a time of arrival at the spacecraft of $\sim$09:53~UTC, which is only $\sim$7~minutes earlier than the observed arrival time of the shock at Solar Orbiter at $\sim$10:00~UTC. Again, the results indicate that our fitting describes reasonably well the shock kinematics. However, \cite{Trotta2024} reported that the shock speed measured in situ at Solar Orbiter was \(\sim 950\) km s\(^{-1}\), presenting a significant discrepancy when compared to the shock speed estimated using the model and ballistic extrapolation. This large difference is unlikely to be explained by inhomogeneities in the medium alone. At such large distances the accuracy of the shock kinematics from the 3D reconstruction is significantly reduced since it relies on only a single viewpoint. Hence, one possibility is that this discrepancy comes from an overestimation of the true shock speed towards this observer. As we explained earlier the two viewpoints do not allow for a tight constraint of the ellipsoid close to the apex. Another plausible explanation lies in the shock’s orientation relative to the ecliptic plane and its propagation upon reaching Solar Orbiter. Observational studies show that the shock was already deviating from the purely radial direction at PSP, with this deviation becoming even more pronounced at Solar Orbiter. Thus, the discrepancy likely arises from a combination of ballistic extrapolation, the shock's propagation geometry, and the non-ideal properties of the medium.

\subsection{Shock Parameters (Mach Numbers and Geometry)}

\begin{figure*}
  \centering
  \includegraphics[width=0.95\textwidth]{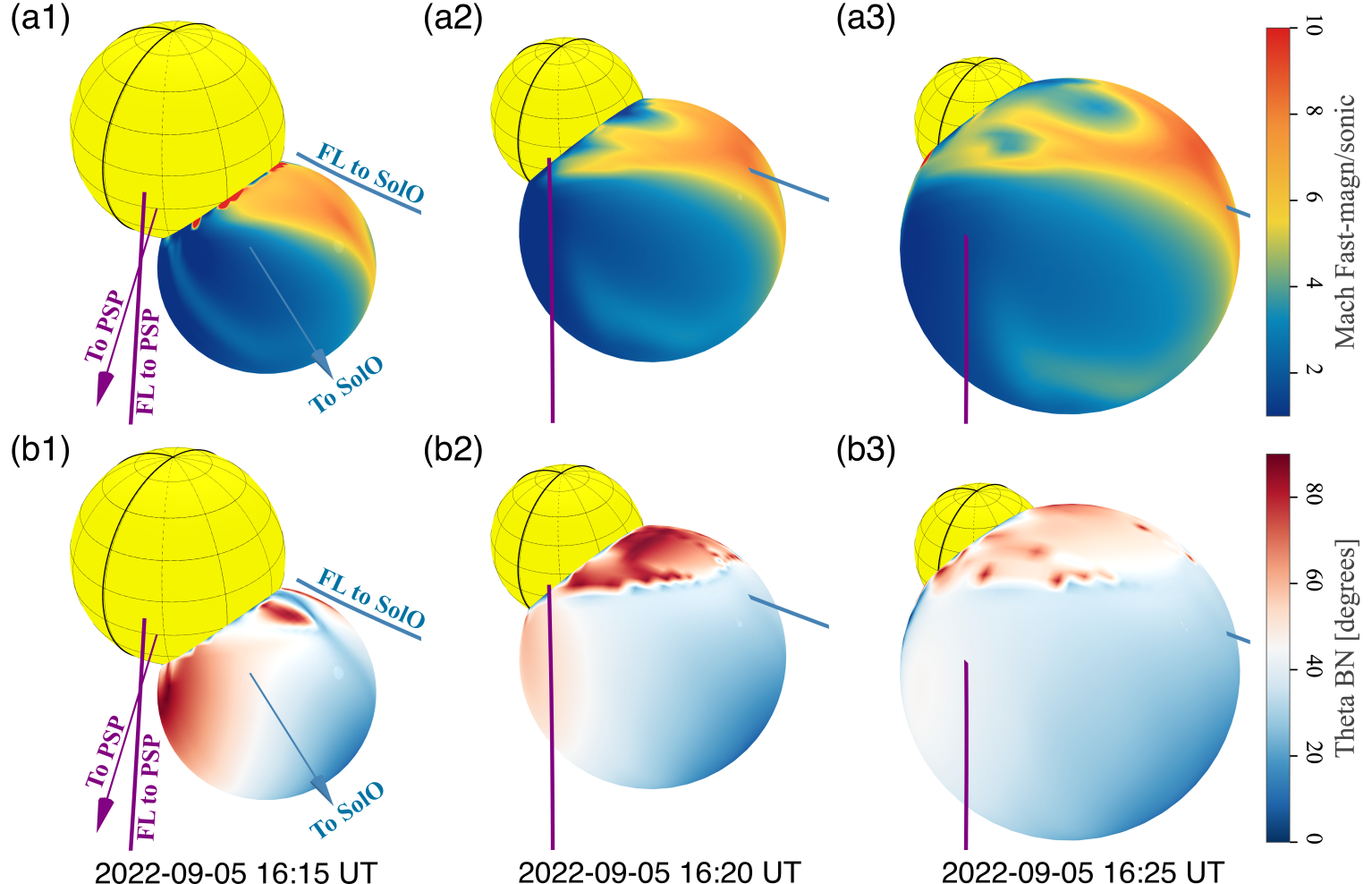}
\caption{The modeled shock wave parameters are plotted in 3D along the reconstructed wavefront surface. The panels from left to right show three selected frames at different times, noted at the bottom. The top row panels (a1 to a3) show the distributions of the modeled fast-magnetosonic Mach number, $M_{fm}$, and the bottom row panels (b1 to b3) the $\Theta_{BN}$ angle (shock geometry). At the first column panels (a1 and b1), the two arrows point towards the two observers and the lines are the magnetic field lines (FL) connecting the two spacecraft, PSP and Solar Orbiter, to the Sun. The Sun (yellow sphere) is plotted to scale at each frame. }
     \label{fig:shock3D_params}
\end{figure*}

For the shock and SEP modelling, we employ data from the MAS thermodynamic model
\citep[see also][]{Riley2019, Riley2021}, which solves the set of resistive MHD equations in spherical coordinates on a non-uniform mesh. This model incorporates plausible energy equations, accounts for radiative losses, and includes a parameterized coronal heating mechanism and considers thermal conduction parallel to the magnetic field \citep{Lionello2009}. By incorporating these detailed thermodynamic effects, the MAS model provides a steady-state estimate of the plasma density and temperature in the corona \citep{Riley2011}. Previous studies have shown that there is a reasonable agreement between the model and observations of the ambient solar wind \citep[e.g.][]{Riley2021}.

The MAS model covers the region from the solar surface to 30 R$_\odot$. For the inner boundary, the model uses a single synoptic map that covers an entire solar rotation, which is constructed by photospheric magnetic field data from the SDO’s Helioseismic and Magnetic Imager instrument \citep{Scherrer2012}. In our study, we specifically used the high-resolution MAS data cube for Carrington rotation 2261 provided by Predictive Science Inc. (https://www.predsci.com/). The resolution of the data cube's mesh for the parameters stored at the cell centers of the control volumes $255 \times 143 \times 300$ in the $r\times\theta\times\phi$ components.

Next, we use the results of the shock kinematics obtained from the 3D reconstruction of the shock, combined with the MHD parameters characterizing the background solar corona, to estimate the shock parameters across the entire shock surface \citep[see][]{Rouillard2016, Kouloumvakos2019}. We compute the 3D expansion speed from the time derivative of the minimum distance between the grid points of the ellipsoid fittings. Subsequently, we calculate the Mach numbers and the magnetic field obliquity with respect to the shock normal, utilizing the plasma and magnetic field properties provided by the MAS model and inferred along the surface of the modeled pressure wave \citep[see details in][]{Rouillard2016, Lario2017, Lario2020, Kouloumvakos2019}.

In Figure~\ref{fig:shock3D_params}, we present a series of selected snapshots displaying the modeled shock wave parameters in 3D, plotted along the reconstructed pressure wavefront surface. Panels~(a) show the fast-magnetosonic Mach number ($M_{fm}$), indicating the presence of multiple regions where a shock wave has formed, both at the apex and flanks of the wavefront. 
Notably, near the field line connecting to Solar Orbiter and mostly over a broad region extending from north to south of the wavefront, there is a specific region that exhibits high Mach numbers (e.g. $\mathrm{M_{fm}\gg}$4), suggesting the formation of strong shock regions in the low corona during the event. 
This specific region is situated in proximity to the heliospheric current sheet, where the local Alfv\'en speed is low because of the high plasma density and weak magnetic field. 
Panels~(b) show the shock obliquity $\Theta_{BN}$, which is the angle between the upstream magnetic field vector and the ellipsoid's surface normal vector. 
The normal can be aligned with or perpendicular to the upstream magnetic field vector, resulting in a $\Theta_{BN}$ angle that varies between 0$^\circ$ to 90$^\circ$. A shock is termed ``parallel" ($\Theta_{BN}=0^\circ$) when the shock normal aligns with the upstream magnetic field, while it is termed ``perpendicular" ($\Theta_{BN}=90^\circ$) when the normal is oriented at a 90-degree angle to the upstream magnetic field. For intermediate $\Theta_{BN}$ angles, the shock can be quasi-perpendicular (quasi-parallel) when $\Theta_{BN}>45^\circ$ ($\Theta_{BN}<45^\circ$). 

As displayed in panels (b) of Figure~\ref{fig:shock3D_params}, we infer that near the shock apex, $\Theta_{BN}$ is predominantly quasi-parallel, while toward the flanks, it is primarily quasi-perpendicular to oblique. Overall, the shock geometry undergoes rapid changes in the low corona due to the shock wave's propagation through coronal regions with differing magnetic topology. These dynamic alterations in the shock geometry are seen when comparing panel (b1) and (b2).

\begin{figure}
  \centering
  \hspace{-1.3em}
  \includegraphics[width=0.49\textwidth]{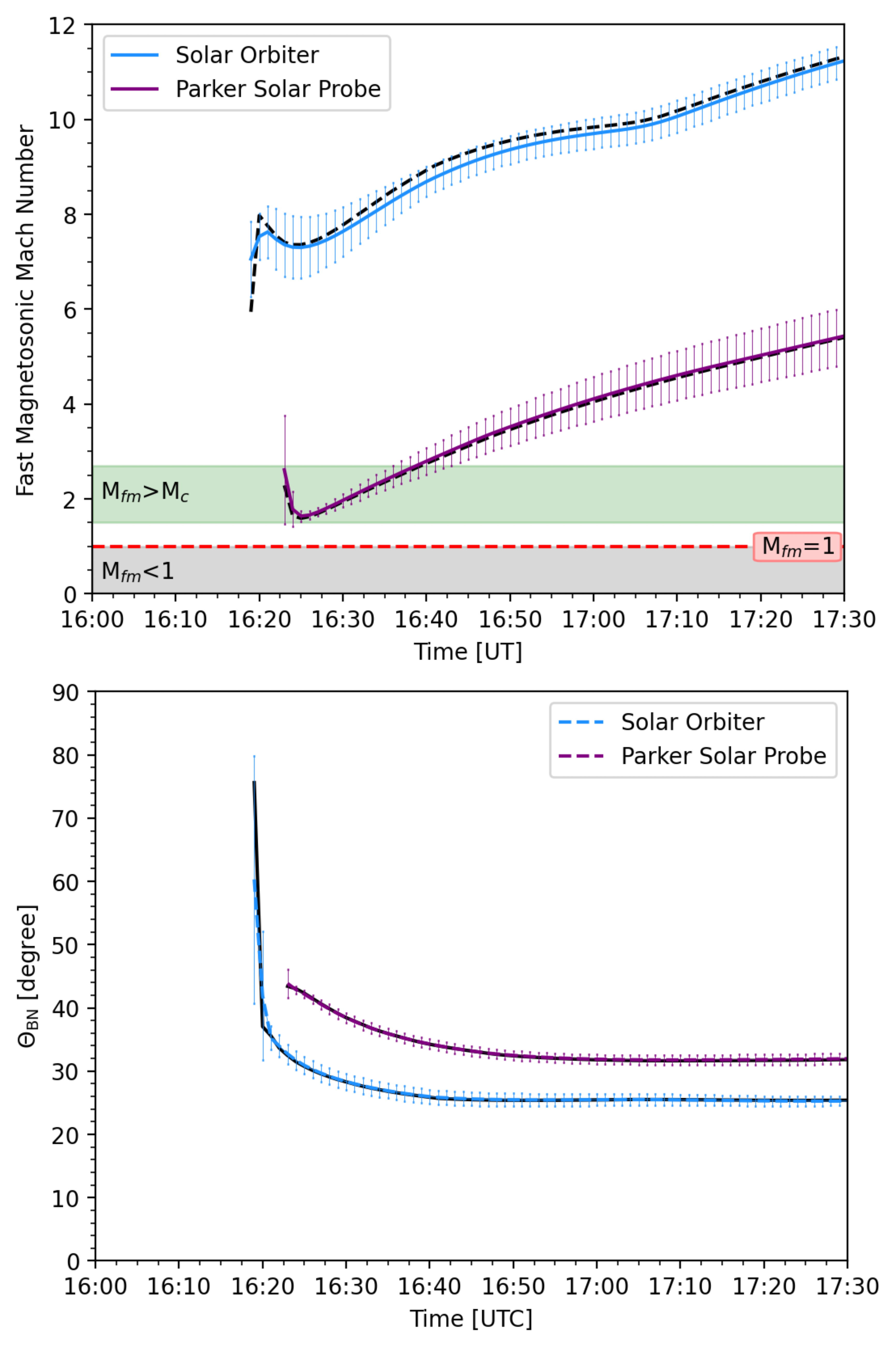}
\caption{ Temporal evolution of the shock $M_{fm}$ (top panel) and $\Theta_{BN}$ angle (bottom panel). The black dashed lines show the shock parameters at the magnetic field lines connecting the two spacecraft to the Sun, and the colored solid lines show the mean values of the shock parameter calculated at a region of 3$^\circ$ around the nominal magnetic connection. The error bars show the standard deviation around the mean. At the top panel, the green shaded area shows the region where the shock becomes super critical ($1.5 \geq M_c \leq 2.7$) and the gray shaded region the region where no shock has formed $M_{\rm fm} \leq 1$.}
     \label{fig:shock3D_params_FLs}
\end{figure}

Figure~\ref{fig:shock3D_params_FLs}, shows the temporal evolution of the shock $M_{fm}$ (top panel) and $\Theta_{BN}$ angle (bottom panel) along the magnetic field lines connecting to the two spacecraft. The uncertainty of the shock parameters has been calculated from the uncertainty in the magnetic connectivity of the two spacecraft. From the temporal evolution of $M_{fm}$ we find that Solar Orbiter is promptly connected to strong shock regions. These regions are supercritical (M$_{fm}>$M$_c$)\footnote{The critical Mach number M$_c$ depends on local plasma conditions (plasma $\beta$) and shock geometry $\Theta_{BN}$. For $\beta\ll1$, the M$_c$ varies from $\sim$1.5 for parallel to $\sim$2.7 for perpendicular shocks \citep[see][for further details]{Laming2022}. } from the beginning of their connection to the spacecraft, with $M_{fm}\sim6.5$. The shock geometry is mainly quasi-perpendicular at the first connection. The strength of the shock regions connected to Solar Orbiter increases rapidly (see Figure~\ref{fig:shock3D_params_FLs}). Within the first hour of the shock evolution, $M_{fm}$ attains a value of 10, reaching its maximum strength at 22:30~UTC with a value of 15.2 (not shown). The $\Theta_{BN}$ angle also evolves rapidly within the first hour and the shock geometry is mainly quasi-parallel reaching asymptotically a $\Theta_{BN}$ value of $\sim$26$^\circ$. For PSP, the magnetic connection with the shock is established approximately 10 minutes later than Solar Orbiter, at 16:21 UTC. Initially, the shock exhibits subcritical behavior with $M_{fm} \approx 1.2$, but its strength gradually increases, reaching around $M_{fm} \approx 5$ by the time the shock reaches the spacecraft. The shock geometry is mainly oblique with $\Theta_{BN} \sim 32^\circ$.

The shock strength and geometry derived from the modeling are not far from the results from in-situ observational studies of the event but do not align precisely either. The \(\Theta_{BN}\) obtained from PSP observations by both \cite{Trotta2024} and \cite{Romeo2023} suggest that the shock was oblique, with \(\Theta_{BN} \sim 42^{\circ}\) and \(\Theta_{BN} \sim 52^{\circ}\), respectively. Our estimations are rather close to the estimate of \cite{Romeo2023}. The Alfv\'en and fast magnetosonic Mach numbers were around \(\sim 4\) according to \cite{Trotta2024}, and \(\sim 2.7\) as estimated by \cite{Romeo2023}. Such discrepancies between observational studies are common due to the challenges of estimating shock parameters using single spacecraft data \citep[see][]{Paschmann00}. Such techniques assume a downstream steady state which is non-trivial to identify in data \citep[][]{gedalin2022,Gedalin22JPP}. While the model parameters are often influenced by the broader inner-heliospheric structure of the magnetized plasma, the local parameters derived from observations are affected by local inhomogeneities and gradients. The variations in shock strength between the two observational studies highlight the inherent challenges in accurately estimating these parameters from in-situ data. Nevertheless, based on both observations, we can conclude that the shock at PSP was oblique and had a moderate Mach number similar to what we find from the shock model. The modeling results for the shock strength deviate from both studies but are closer to the estimate provided by \cite{Trotta2024} (by $\sim$20\%). As we explained earlier, one possibility is that an overestimation of the shock speed would lead to somewhat higher shock strength, while also, inhomogeneities in the medium can affect any comparison between a global model and in situ measurements.

\subsection{SEP Transport model}\label{sec:paradise}

This study utilizes the model developed by \cite{Wijsen2020} to track the evolution of SEPs through the solar corona and inner heliosphere. The PARADISE model solves the five-dimensional focused transport equation \citep[see][for a recent review]{vandenBerg2020}, which characterizes the time-dependent evolution of an energetic particle distribution propagating through a denser background plasma. In our PARADISE simulations, we employ the MAS simulations of the corona to provide a description of the background plasma through which the SEP distributions evolve.

The focused transport equation solved by PARADISE incorporates the effects of solar wind turbulence described as a diffusion process in the particles' pitch-angle and a spatial diffusion process perpendicular to the average interplanetary magnetic field. In this study, we adopt the results of quasi-linear theory \citep[][]{jokipii66} for the pitch-angle diffusion coefficient and the non-linear guiding center theory \citep{matthaeus2003} to describe the cross-field diffusion process. For a detailed overview of the focused transport equation, the functional forms of these diffusion coefficients and the underlying turbulence parameters, we refer the reader to Appendix~\ref{app:paradise}.

In this work, we do not model the actual particle acceleration at the shock with PARADISE, but instead rely on the results of diffusive shock acceleration to prescribe an accelerated particle distribution at the shock. This particle distribution depends on the local shock parameters and varies thus both in time and space.
In particular, we assume that the particle distribution at the shock is given by

\begin{equation}\label{eq:DSA}
    f_\mathrm{s}(p) = \frac{\sigma \epsilon n_\mathrm{p}}{4\pi p_\mathrm{inj}^3}\left(\frac{p}{p_\mathrm{inj}}\right)^{-\sigma}\exp\left[-\left( \frac{p}{p_c}\right)^2 \right],
\end{equation}

\noindent where \(\sigma\) is the spectral index, \(n_\mathrm{p}\) is the upstream proton number density, and \(\epsilon = 5\times10^{-5}\) is the fraction of protons assumed to be injected into the shock acceleration mechanism at the injection momentum \(p_\mathrm{inj}\) \citep{Afanasiev2024} . The parameter \(p_c\) represents the roll-over momentum.  Following standard diffusive shock acceleration, the spectral index can be written in terms of the scattering-center compression ratio $r_c$ across the shock as $\sigma = 3 r_c/(r_c - 1)$. 
Here, $r_\mathrm{c} = r_\mathrm{g}(1 - M_\mathrm{A}^{-1})$, with
$r_\mathrm{g}$ representing the gas compression ratio across the shock wave and $M_\mathrm{A}$ is the Alfvenic Mach number (defined in the de Hoffmann-Teller frame). Hence, $r_c$ represents the compression ratio of the scattering centers, assuming that the scattering centers in the upstream region are Alfv\'en waves and that in the downstream region, they are magnetic fluctuations advected with the flow. 

We further assume that the injection momentum in Eq.~\eqref{eq:DSA} is given by $p_\mathrm{inj} = 2m_p\Delta u$, with $\Delta u$ denoting the jump in the solar wind speed across the shock wave. This assumption is based on the scenario in which upstream solar wind protons experience head-on collisions with the downstream scattering centers, which are assumed to advect with the flow \citep[e.g.,][]{Vainio2014}. 
It is worth noting that in reality, the injection momentum could be influenced by additional factors such as the shock obliquity and the diffusion conditions of particles in both perpendicular and parallel directions to the background magnetic field, among other factors. However, since the diffusion conditions of energetic particles in the vicinity of the shocks are not well understood, we opt for a simpler approach. 

\begin{figure}
    \centering
    \includegraphics[width=0.45\textwidth]{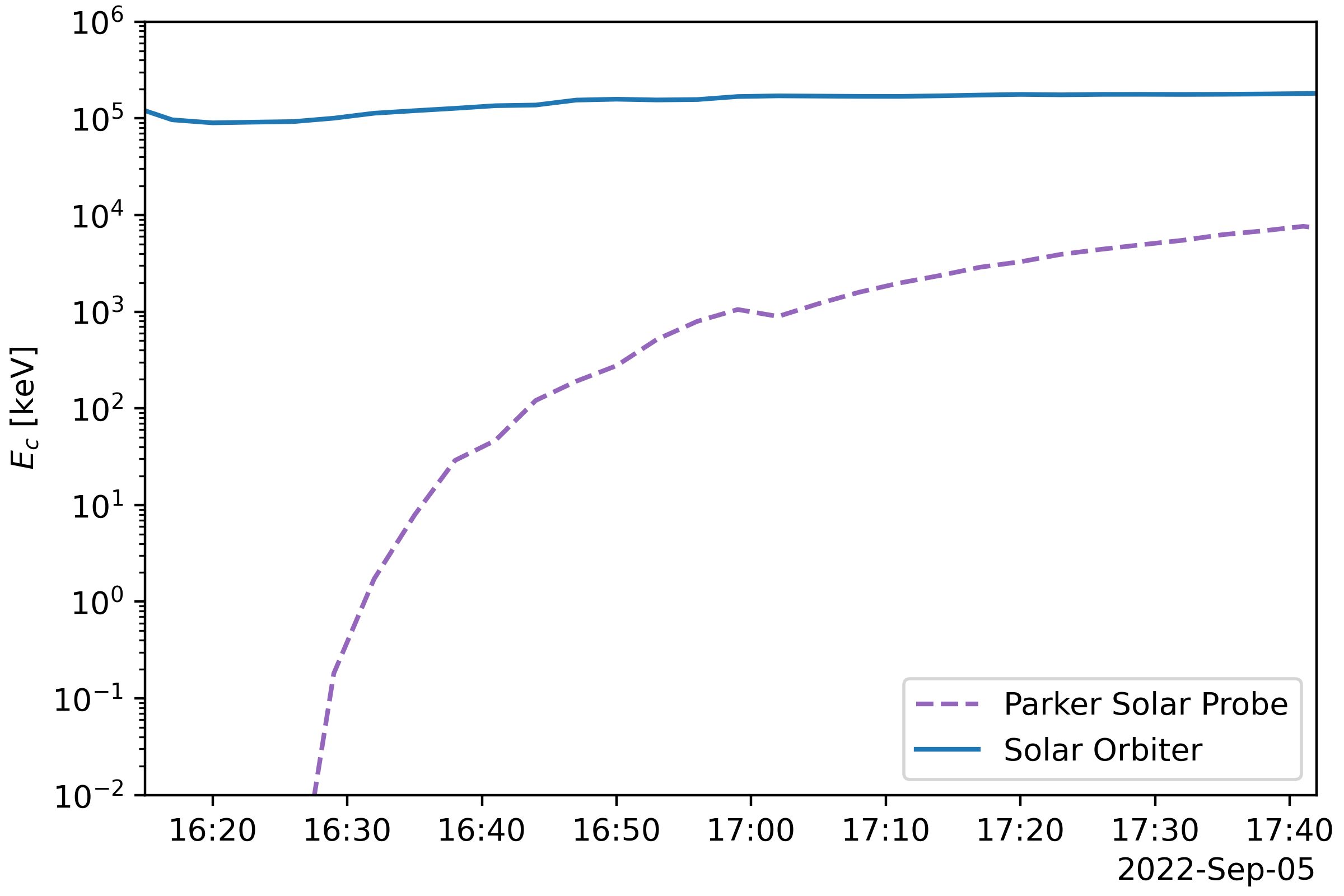}
    \caption{The roll-over energy of the particle spectrum at the shock points that connect to the Solar Orbiter and PSP. }
    \label{fig:roll-over-energy}
\end{figure}

The exponential term in Eq.~\eqref{eq:DSA} is included to account for the effects of finite acceleration time and a foreshock with finite extent. We prescribe the roll-over momentum $p_c$ by using equation~(18) of \citet{Vainio2014}, that is,
\begin{equation}\label{eq:pc}
    p_c= \eta p_\mathrm{inj}\left(\frac{\pi\epsilon\sigma}{4} \frac{r_s}{d_i}\right)^{1/({\sigma - 3} )},
\end{equation}

\noindent with $d_i = V_A/\Omega_p$ the proton inertial length and $\Omega_p$ the proton gyrofrequency. Taking into account the dependence on the radial coordinate \( r_s \), this equation considers that a point on the shock surface has had more time to accelerate particles to higher energies as it reaches greater radial distances.
This roll-over momentum was derived by \citet{Vainio2014}, considering that the turbulent trapping of particles in the foreshock loses effectiveness beyond a certain distance from the shock due to the effect of adiabatic focusing. Moreover, as explained in \citet{Vainio2014}, Eq.~\eqref{eq:pc} overestimates the roll-over momentum, as it is derived assuming a  diffusion coefficient in the foreshock that is constant in space and time. To account for this, we have introduced the reduction factor $\eta$ in Eq.~\eqref{eq:pc}, which we set equal to 0.35. The latter value gives a good match with PSP observations in the simulations presented below. 

Figure~\ref{fig:roll-over-energy} shows the roll-over energy $E_c$ corresponding to the roll-over momentum given in Eq.~\eqref{eq:pc}. It can be seen that the roll-over energy of Solar Orbiter is high from the onset ($\sim 100$ MeV) of the simulation and increases slowly. This high onset results from the fact that Solar Orbiter is from the very beginning connected to a very strong part of the shock wave. In contrast, PSP is connected to the weaker east flank of the shock, resulting in a very low initial roll-over energy. Over time, PSP's connection point moves to stronger parts of the shock at larger radial distances, which have thus had more time to accelerate particles. As a result, the roll-over energy increases and reaches ${\sim}10$~MeV by the time the shock crosses the spacecraft.

\begin{figure*}
  \centering
  \includegraphics[width=0.35\textwidth]{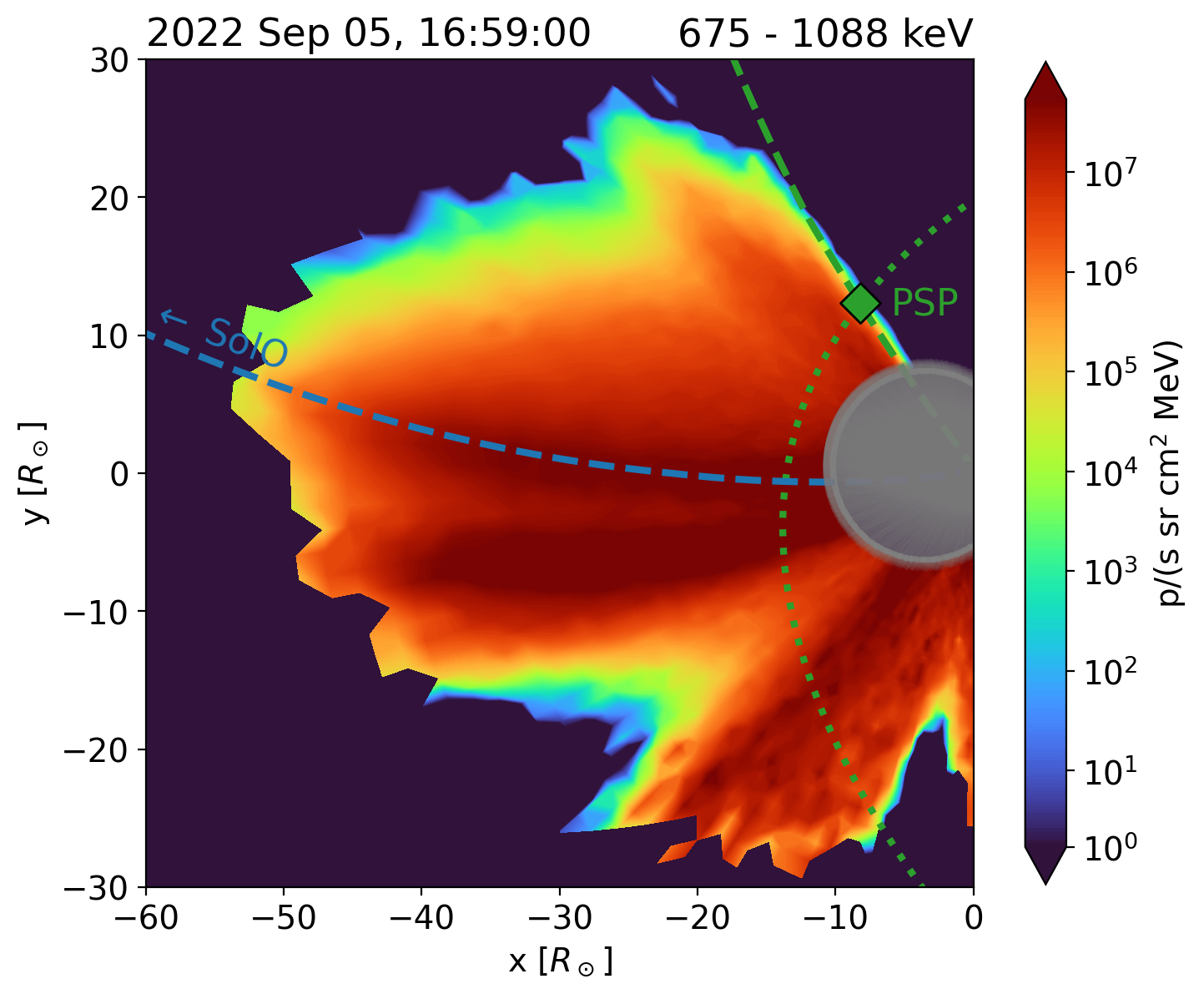}
  \includegraphics[width=0.35\textwidth]{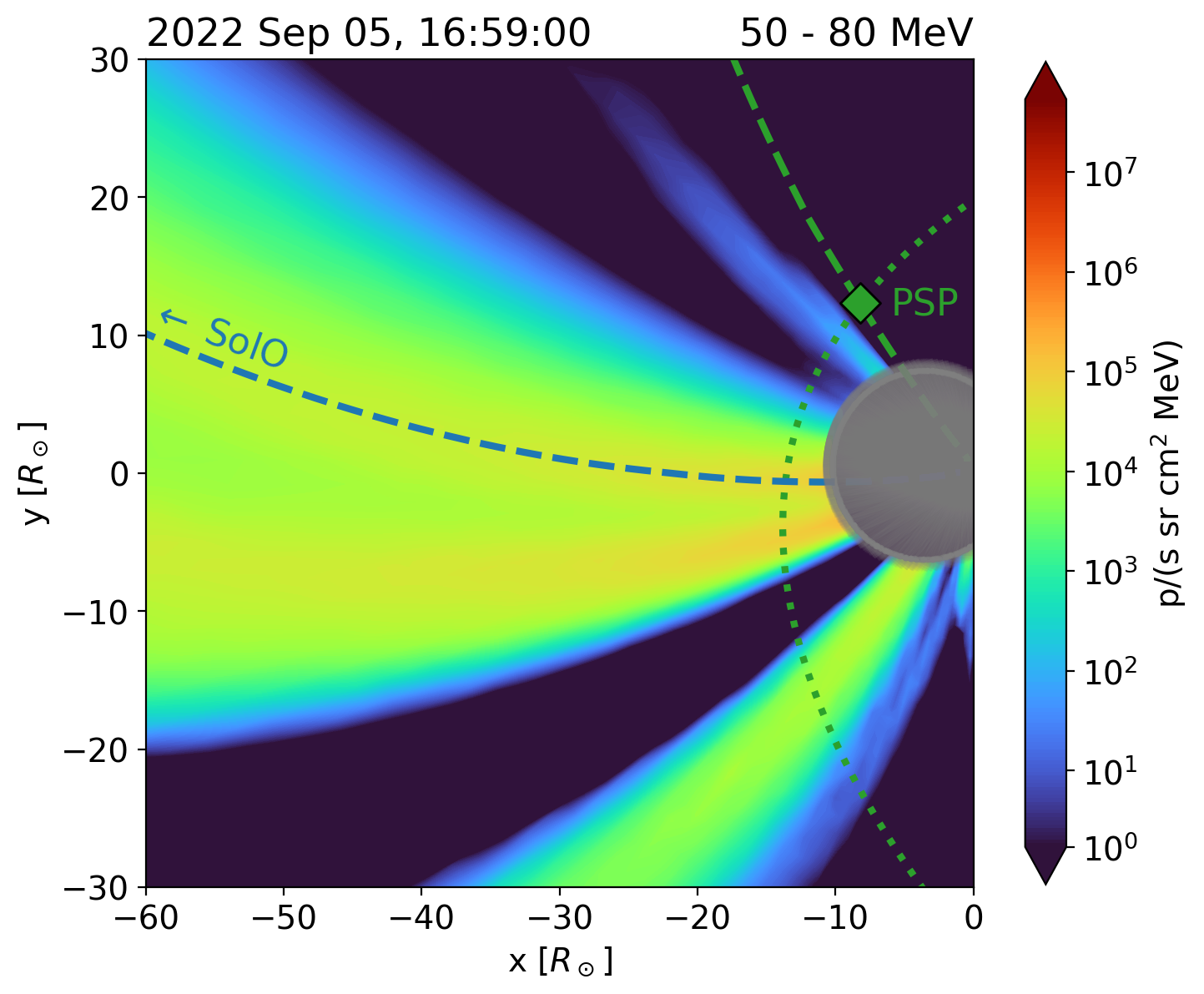}
  \includegraphics[width=0.35\textwidth]{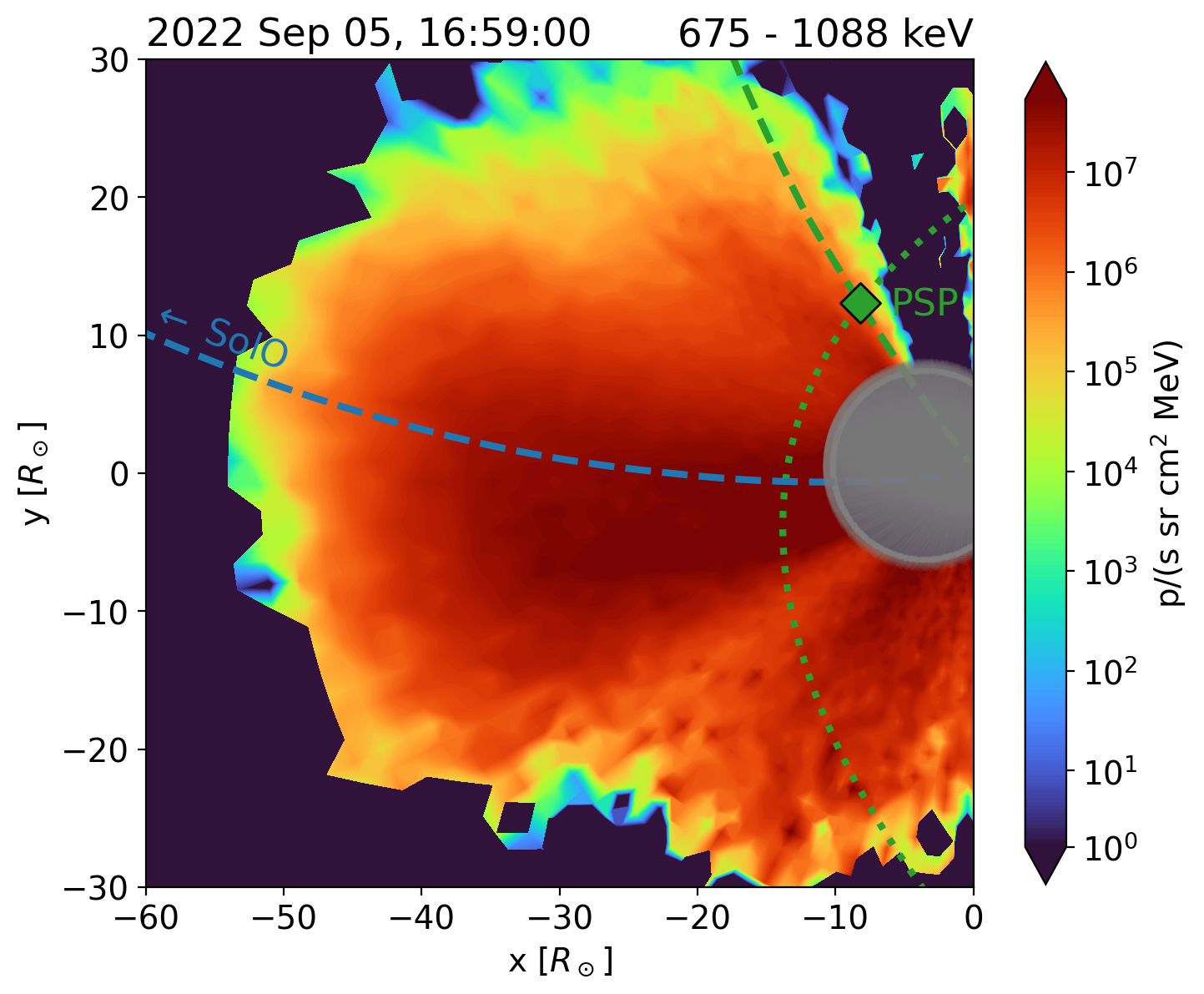}
  \includegraphics[width=0.35\textwidth]{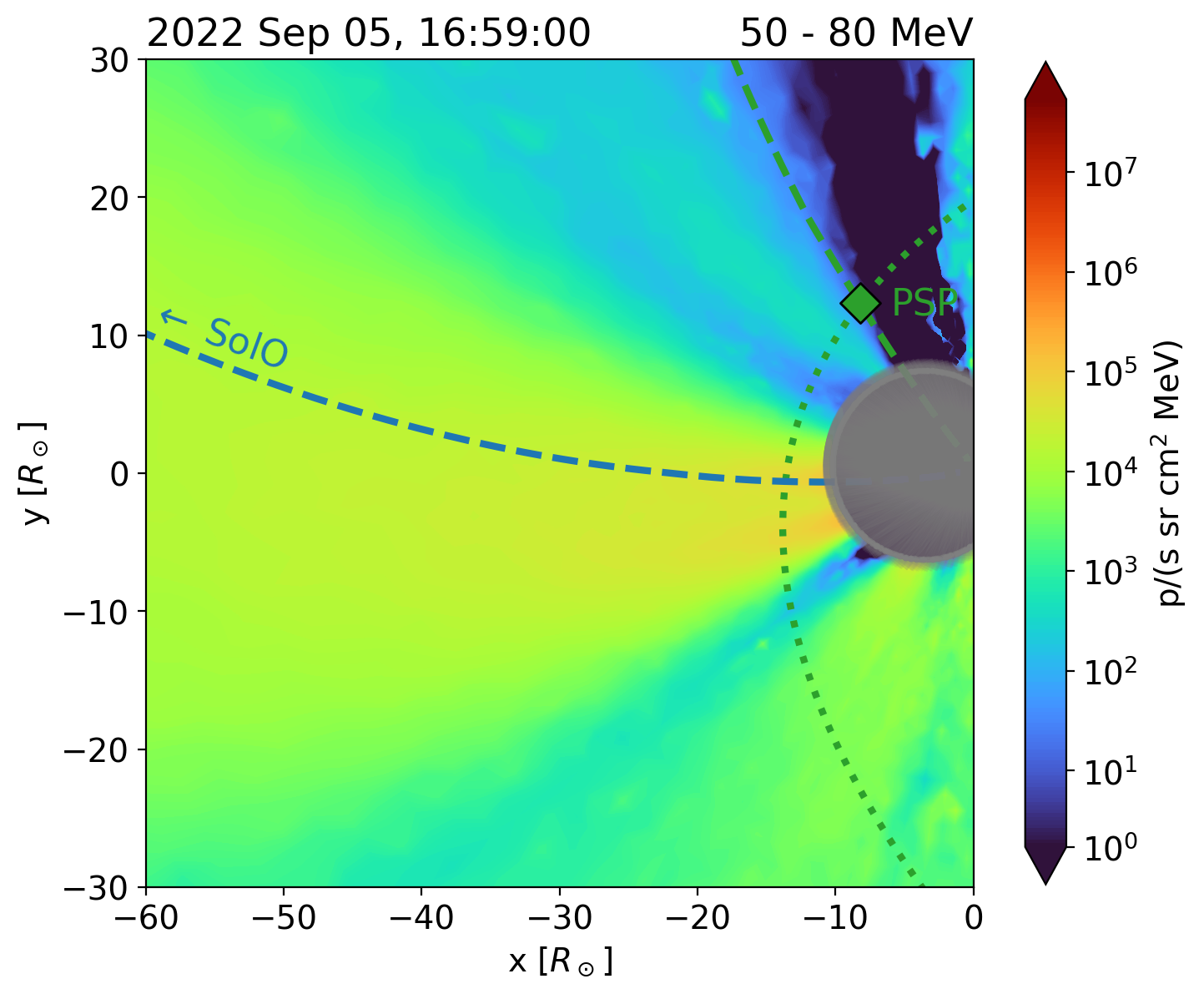}
\caption{A snapshot from the northern ecliptic pole from the PARADISE Model on 2022 Sep 5, 16:59UT. Left panels show omnidirectional proton intensities for energies 675~keV to 1~MeV), while right panels show the same for energies 50\,--\,80~MeV. Top row: the model results with no cross-field diffusion; bottom row: with cross-field diffusion. The gray surface denotes the shock front, while the dashed blue and green lines represent the magnetic field lines connecting to Solar Orbiter and PSP, respectively. PSP's orbit is marked by the green dotted line. An animation of this figure is available. The animation begins at 16:11UT and ends at 17:27UT, and shows the SEP modeling results from the PARADISE Model with (top panels) and without  (bottom panels) cross-field diffusion. 
}
     \label{fig:SEPs_model_distr}
\end{figure*}

\begin{figure*}
  \centering
  \includegraphics[width=0.45\textwidth]{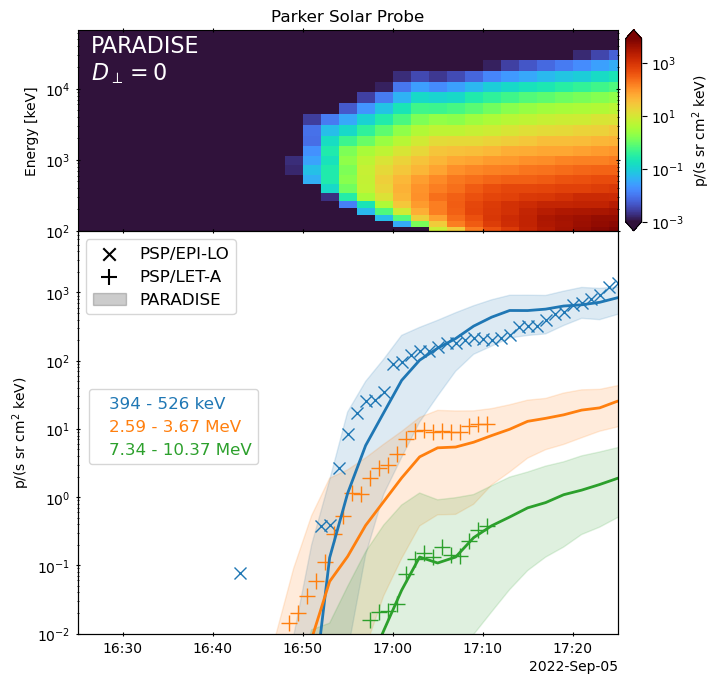}
  \includegraphics[width=0.45\textwidth]{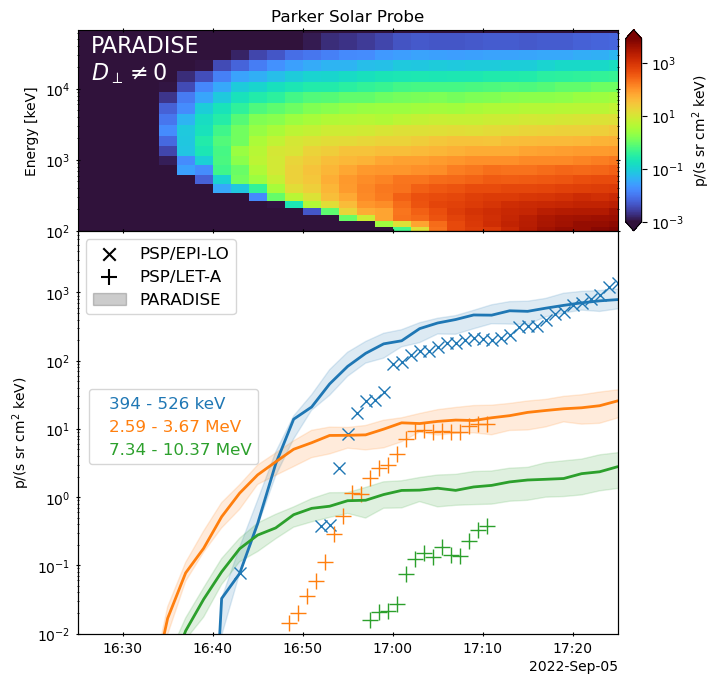}
\caption{Modeled omnidirectional particle intensities at PSP. The left and right figures are for simulations without and with cross-field diffusion, respectively. The top panels show particle spectrograms, while the lower panels show time-intensity profiles for a selection of energy channels. In these profiles, the shaded area indicates the variation of intensities for observers positioned 3$^\circ$ in longitude ahead and behind PSP. Finally, symbols indicate the intensities measured by EPI-LO (crosses) and LET-A (plus signs).  
}
     \label{fig:psp_sims}
\end{figure*}

\begin{table}
\centering
\caption{Time difference, $\Delta t = t_{\text{obs}} - t_{\text{sim}}$, between the observed and simulated times when the energy channels exceed a threshold of $10^{-2}$ p/(cm$^2$ sr s keV). The third and fourth columns present the results for the simulations without and with cross-field diffusion, respectively, for different energy channels.}

\begin{tabular}{c c c c}
\tableline
 & Energy Channel & $\Delta t$ [min]  & $\Delta t$ [min]  \\
 && $D_\perp = 0$ & $D_\perp \neq 0 $ \\
\hline
\hline
\multirow{12}{*}{\rotatebox{90}{EPI-Lo}}
& 123 -- 149 keV & -4.97 & -2.97 \\
& 149 -- 184 keV & -2.97 & 1.03 \\
& 184 -- 232 keV & -2.97 & 1.03 \\
& 232 -- 299 keV & 1.03 & 7.03 \\
& 299 -- 394 keV & 0.03 & 10.03 \\
& 394 -- 526 keV & 0.03 & 12.03 \\
& 526 -- 711 keV & 1.03 & 11.03 \\
& 711 -- 962 keV & 0.03 & 14.03 \\
& 0.96 -- 1.31 MeV & 0.03 & 14.03 \\
& 1.31 -- 1.79 MeV & 2.03 & 16.03 \\
& 1.79 -- 2.59 MeV & 1.03 & 17.03 \\
\hline
\multirow{8}{*}{\rotatebox{90}{LET-A}}
& 2.59 -- 3.67 MeV & -3.47 & 14.53 \\
& 3.67 -- 5.19 MeV & 0.53 & 16.53 \\
& 5.19 -- 7.34 MeV & -1.47 & 18.53 \\
& 7.34 -- 10.37 MeV & -0.47 & 21.53 \\
& 10.37 -- 14.67 MeV & -0.47 & 23.53 \\
& 14.67 -- 20.75 MeV & -1.47 & 28.53 \\

\hline
\end{tabular}
\label{tab:onset_times}
\end{table}

Figure~\ref{fig:SEPs_model_distr} presents selected snapshots from the PARADISE model, illustrating particle intensities for low-energy (675~keV - 1~MeV) protons (left column) and high-energy (50\,--\,80~MeV) protons (right column) in the plane containing PSP orbit. The gray ellipse denotes the shock location, while the dashed lines represent magnetic field lines connecting  PSP (green) and Solar Orbiter (blue) to the shock. 
The trajectory which PSP follows (in counter-clockwise direction) is depicted by the green dotted line. The top row displays simulations without cross-field diffusion ($D_\perp = 0$), while the bottom row shows results incorporating cross-field diffusion (as described in Eq.~\eqref{eq:nlgc}). 
From this figure and the accompanying movie, it appears that there is a stream with elevated particle intensities directed toward Solar Orbiter. This is consistent with what we showed in the previous section, indicating that this spacecraft established magnetic connection with the shock close to its apex, where the shock has a high Mach number. Our model predicts a very efficient acceleration of protons in these regions, as also reflected by the high roll-over energy shown in Fig.~\ref{fig:roll-over-energy}.

In contrast, PSP is located in a region characterized by significantly lower particle intensities, attributed to the weak shock formation at the east flank of the expanding CME. Particularly, the very east flank of the shock is completely devoid of SEPs, as the roll-over energy in this flank is well below the two depicted energy channels in Figure~\ref{fig:psp_sims} and because we only inject particles where the ellipsoid represents an actual shock ($M_{\rm fms} > 1$). As a result, PSP initially observes predominantly low-energy SEPs in the simulation.

However, the expansion of the shock wave and the rapid motion of PSP through the corona propel the spacecraft into a region with elevated SEP intensities at increasingly high energy. Consequently, a significant outcome of our simulation is that, for this event, the observed SEP intensities by PSP may stem from the spacecraft's rapid motion, effectively sampling SEP intensities originating from a shock wave exhibiting significant variations in strength across its surface.

As expected, the incorporation of cross-field diffusion (lower panels in Figure \ref{fig:SEPs_model_distr}) leads to a reduction in the spatial gradients observed in the intensity profiles. This effect is especially noticeable in the vicinity of PSP, as the spacecraft is situated in a region with a steep spatial gradient of particle intensities (as observed in scenarios without cross-field diffusion).

As a next step of this analysis, we compare the results derived from the SEP simulation and the observations recorded by PSP, focusing on EPI-Lo measurements. The top panels of Figure~\ref{fig:psp_sims} show energy spectrograms from the modeled omnidirectional particle intensities at PSP. A first remark is that both simulations, with and without cross-field diffusion, reproduce the inverse velocity dispersion that was observed in EPI-Lo data at energies above $\sim$1 MeV. This is evident in both spectrograms, where the first arriving low-energy protons exhibit a typical velocity dispersion: higher energetic particles arrive earlier than the slower, less energetic ones. By contrast, above about $\sim$1 MeV, the energetic protons show an inverse velocity dispersion. In the simulation, this behavior is a direct consequence of the shock being weak on the east flank, requiring time to accelerate particles to high energies, thus gradually increasing the roll-over energy $E_c$ (see Fig.~\ref{fig:roll-over-energy}).

Furthermore, the simulation without cross-field diffusion appears to reproduce qualitatively better the observed dispersion of SEPs (normal and inverse). Specifically, the inverse velocity dispersion at the modeled spectrogram forms at energies above $\sim$1~MeV, consistent with observations made by EPI-Lo. Conversely, in the simulation with cross-field diffusion, we observe that the inverse velocity dispersion forms at energies above $\sim$3~MeV, which is inconsistent with the observations. However, considering the uncertainties of the shock modelling it is difficult to conclude if this difference in the transition energy where the inverse velocity dispersion forms is significant or not. Therefore, we can only rely on the timing comparison between the two models which shows a significant worsening of the agreement between the SEP simulation and observations as we will show later.

In the bottom panels of Figure~\ref{fig:psp_sims}, we present a comparison between the modeled omnidirectional particle intensities and the time-intensity profiles for selected energy channels of EPI-Lo and EPI-Hi. For the simulation without the cross-field diffusion, we observe a reasonable match between the modeled (solid lines) time profiles and the observed data from EPI-Lo (symbols). The modeled particle intensities exhibit a similar magnitude to the observed ones, with most energy channels falling within the shaded areas, representing the variation of simulated SEP intensities derived from a $\pm$3 degrees change in the longitude of PSP.

Furthermore, the onset times of the modeled SEP profiles align well with the observations for most of the selected energy channels. However, with cross-field diffusion, we note significant discrepancies, primarily in the timings between the modeled and the observed proton intensities. For example, the time difference between the modeled and the observed onset time for $\sim$3~MeV protons is more than 15 minutes. Summarizing the above results we have that the simulation without the cross-field diffusion captures well the timing and the intensity of the event, whereas the simulation with cross-field diffusion fails to reproduce the true onset of the energetic particles for all energies despite the reasonable match between the modeled and the observed intensities (when time-shifted) for some energies.

While only three energy channels are displayed in Figure~\ref{fig:psp_sims} to prevent overcrowding, Table \ref{tab:onset_times} provides a more comprehensive comparison of the timing differences across a broader range of energy channels. The table presents the time difference, \(\Delta t = t_{\text{obs}} - t_{\text{sim}}\), between the observed and simulated times for both simulations (with and without cross-field diffusion), when the particle intensities exceed a threshold of \(10^{-2}\) p/(cm$^2$ sr s keV). A negative \(\Delta t\) indicates that the simulated time at which the threshold is exceeded occurs after the observed time. We adopt this threshold instead of, for example, the exact onset time because the onset time in a stochastic code is not well-defined due to low initial statistics and can be affected by superluminal propagation \citep{Strauss2023b}. The results in Table \ref{tab:onset_times} indicate that the simulation without cross-field diffusion consistently provides a closer match to the observed timing, as \(\Delta t\) is generally closer to zero. Only at the lowest simulated energies ($<200$ keV), the simulation with cross-field diffusion performs slightly better.

\begin{figure}
  \centering
  \includegraphics[width=0.48\textwidth]{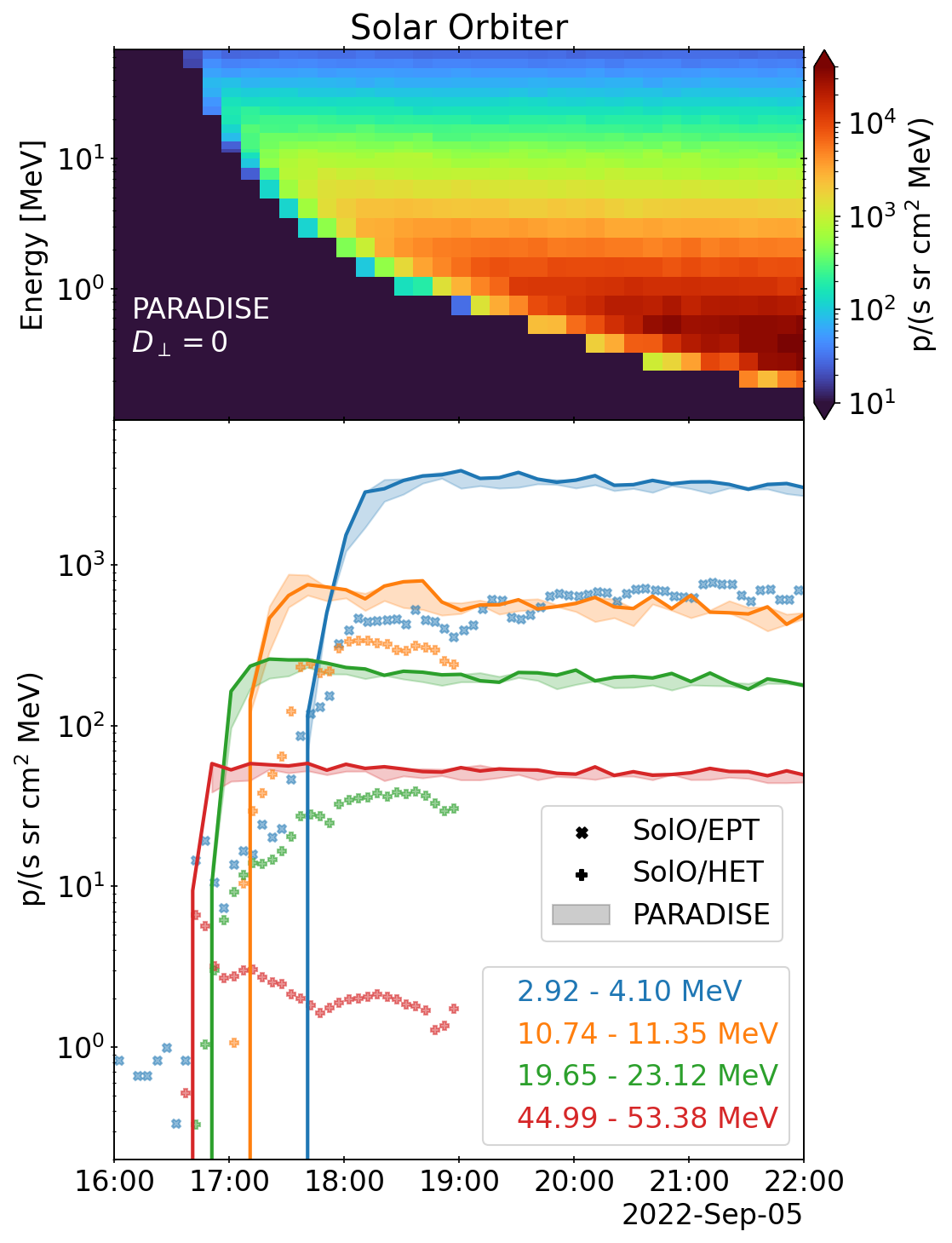}
\caption{Modeled omnidirectional particle intensities at Solar Orbiter. The top panel shows the particle spectrogram, while the lower panel shows time-intensity profiles for a selection of energy channels. In these profiles, the shaded area indicates the variation of intensities for observers positioned 3 degrees in longitude ahead and behind Solar Orbiter. Finally, markers indicate the intensities measured by EPT (crosses) and HET(plus signs).   }
     \label{fig:Solar Orbiter_sim}
\end{figure}

Figure~\ref{fig:Solar Orbiter_sim} shows the PARADISE simulation results obtained at Solar Orbiter. We only  show the simulation without cross-field diffusion, as the one with cross-field diffusion does not show notable differences. This is a consequence of Solar Orbiter being magnetically connected to the shock apex, where the particle acceleration efficiency is the highest in the simulation. Observations of near-Sun strong shocks have also revealed a complex but efficient acceleration process taking place at the apex, where the shock is predominantly quasi-parallel \citep[][]{Jebaraj2024}. The upper panel of Figure~\ref{fig:Solar Orbiter_sim} illustrates the energy spectrogram generated from the modeled  particle intensities at Solar Orbiter. In agreement with the observations, the simulations at Solar Orbiter display a conventional velocity dispersion pattern, where the most energetic particles reach the spacecraft first (see also Figure~\ref{fig:SEP_Observations}).

In the lower panel of Figure~\ref{fig:Solar Orbiter_sim}, we present a comparison between the modeled and observed particle time-intensity profiles. We note that agreement between simulated and observed onset times improves at higher particle energies. However, the model shows a steeper rise at the onset of the SEP event compared to the observations; in the observations, intensities peak only approximately one hour after the event onset. This delay in the observed peak coincides with an increase in the non-radial magnetic field components at Solar Orbiter (not shown) and may thus result from a change in magnetic field connectivity that is not captured by the model.

The overall match between the simulations and the observations at Solar Orbiter is not as close as at PSP. Specifically, the simulated intensities at Solar Orbiter are roughly an order of magnitude too high and exhibit a significantly harder energy spectrum than the observations. This discrepancy suggests that the injection and roll-over momentum, as given in Eq.\eqref{eq:pc}, is substantially overestimated at the shock apex points connected magnetically to Solar Orbiter. This is also reflected in Figure~\ref{fig:roll-over-energy}, which shows that the roll-over energy at the shock point connected to Solar Orbiter is already high (at around 100 MeV) from the start of the simulation. This may be a consequence of an overestimation of the shock strength at its apex. In addition,  our adoption of a somewhat larger reduction value (see Eq.~\eqref{eq:pc}), $\eta = 0.35$ , was motivated by the aim to optimize agreement with the PSP observations and reflects the weak shock conditions encountered by PSP. Nonetheless, it is expected that $\eta$ varies across the shock surface, as this parameter serves as an indicator of how rapidly the diffusion coefficient in the foreshock reaches a steady-state value \citep[see][for further details]{Vainio2014}.

\section{Discussion and Conclusion}

In this study, we analyzed a significant SEP event on September 5, 2022, observed by both PSP at a heliocentric distance of 15.4~R$_\sun$ and Solar Orbiter at 151.1~R$_\sun$. VDA of the SEP measurements revealed that SEPs were released promptly at Solar Orbiter but experienced a substantial delay at PSP. Additionally, the SEP travel path at Solar Orbiter aligned with the Parker spiral model \citep[e.g.][]{Vainio2013JSWSC}, while at PSP, the path length was much shorter, suggesting that SEPs were released higher in the corona \citep[e.g.][]{Zhu2018, Kouloumvakos2022}.

Our observations indicate that at Solar Orbiter, SEPs were released almost simultaneously with the shock's first contact with the magnetic field lines, consistent with strong shock regions connected to the spacecraft. In contrast, PSP was initially connected to a weaker, subcritical shock region, resulting in delayed SEP release as the shock strengthened and became super-critical. This delay aligns with previous studies suggesting that shock evolution in the corona significantly affects SEP release times \citep[e.g.][]{Kouloumvakos2022, Kouloumvakos2023, Zhuang2024}.

To understand these observations, we employed the PARADISE model alongside 3D shock modeling. The model successfully reproduced the SEP onset and inverse velocity dispersion observed by PSP, supporting the idea that SEP acceleration is linked to shock strength and geometry. However, discrepancies arose when comparing model predictions with Solar Orbiter data, likely due to oversimplified assumptions in particle injection efficiency and shock parameter uniformity in the model.

Furthermore, our simulations showed that including cross-field diffusion reduced the agreement with PSP observations, suggesting that perpendicular transport may play a minor role in SEP transport for this event, at least at the close distances where PSP was located. This finding challenges some recent studies \citep{Strauss2023, Posner2024} and indicates that transport conditions may vary significantly from event to event.

The main results of the study may therefore be summarized into the following four points.

\begin{itemize}
    \item SEPs at PSP were released with a significant delay and followed a shorter path than the Parker spiral, indicating that the release of SEPs occurred higher in the low corona \citep[e.g.][]{Jebaraj23L}.
    
    \item The shock's characteristics at PSP, initially subcritical and later super-critical, were crucial in the delayed SEP release and the inverse velocity dispersion observed above $\sim$1 MeV at this spacecraft. This behavior is consistent with the shock dynamics influencing SEP release times \citep[e.g.][]{Kouloumvakos2022}.
    
    \item The PARADISE model effectively reproduced PSP observations but overestimated SEP intensities and significantly harder energy spectrum at Solar Orbiter, highlighting the need for more nuanced treatment of the efficiency of particle injection  \citep{Vainio2014} and further advancements in the modeling of shock parameters.
    
    \item Cross-field diffusion appeared negligible in SEP transport to PSP, suggesting a variability in transport mechanisms.
\end{itemize}

This study highlights the value of SEP measurements close to the Sun for unraveling the early phases of SEP acceleration and transport. The contrasting behaviors observed by PSP and Solar Orbiter suggest a complex interplay between shock dynamics and particle acceleration. However, other processes could also have contributed to the observed differences. The development of more comprehensive models that integrate various acceleration and transport mechanisms is needed to evaluate the interplay among the different physical processes. Moreover, future work should focus on refining models to account for varying shock properties and exploring additional events to better understand SEP release mechanisms. 

Finally, our study emphasized the importance of combined shock and SEP modeling, which clarified some aspects of the observed SEP properties. For SEP modeling, we employed a relatively simple semi-analytical model for shock acceleration based on steady-state diffusive shock acceleration \citep{Bell1978}, with modifications to account for finite-time effects \citep{Vainio2014}. Notably, we did not include a foreshock region by prescribing a small mean free path near the shock \citep{Wijsen2022}. Including such features would be necessary to improve agreement between observations and simulations near the shock. The satisfactory agreement between PSP observations and our simulation is likely due to the limited spatial extent of the foreshock, a consequence of the shock being weak in the flank encountered by PSP. However, more advanced models, such as Particle-in-Cell models, are required to fully capture the ongoing processes of particle acceleration at shock waves \citep{Guo2013, Caprioli2014}.

\vspace{1em}
\section*{Acknowledgments}

The authors would like to thank the anonymous reviewer for their thoughtful and constructive comments which helped to improve the quality of this paper.
A.K.\ acknowledges financial support from NASA NNN06AA01C (PSP EPI-Lo) contract.
N.W.\ acknowledges funding from the Research Foundation -- Flanders (FWO -- Vlaanderen, fellowship no.\ 1184319N) and from the KU Leuven project 3E241013.
A.A.\ acknowledges funding from the European Union’s Horizon 2020 research and innovation programme under grant agreement No 101004159 (SERPENTINE) as well as from the Finnish Centre of Excellence in Research of Sustainable Space (FORESAIL) funded by the Research Council of Finland (grant no.\ 336809).
I.C.J is grateful for support from the Research Council of Finland (SHOCKSEE, grant No.\ 346902) and the European Union’s (E.U's) Horizon 2020 research and innovation program under grant agreement No.\ 101134999 (SOLER). The study reflects only the authors' view and the European Commission is not responsible for any use that may be made of the information it contains.
Computational resources and services used in this work were provided by the VSC (Flemish Supercomputer Centre), funded by the FWO and the Flemish Government-Department EWI. 
Parker Solar Probe was designed, built, and is now operated by the Johns Hopkins Applied Physics Laboratory (JHU/APL) as part of NASA’s Living with a Star (LWS) program (contract NNN06AA01C). We thank the STEREO: SECCHI SOHO: LASCO; SDO/AIA teams and Predictive Science Inc. for providing the data used in this study. The STEREO SECCHI data are produced by a consortium of RAL (UK), NRL (USA), LMSAL (USA), GSFC (USA), MPS (Germany), CSL (Belgium), IOTA (France) and IAS (France). SOHO is a mission of international cooperation between ESA and NASA. The SDO/AIA data used are courtesy of SDO (NASA) and the AIA consortium. Some data processing for this research was carried out using version 5.1.0 (Mumford et al. 2021) of the SunPy open source software package (SunPy Community 2020). This research has made use of PyThea v0.11.0, an open-source and free Python package to reconstruct the 3D structure of CMEs and shock waves (Zenodo: https://doi.org/10.5281/zenodo.5713659).

\newpage
\bibliography{sample631}{}
\bibliographystyle{aasjournal}

\appendix

\section{PARADISE set-up}\label{app:paradise}

The transport equation solved by the PARADISE model is the so-called  focused transport equation \citep[e.g.,][]{skilling71, isenberg97, leRoux09,vandenBerg2020}, which is given by

\begin{multline} \label{eq:fte}
\pd{j}{t} + \nabla\cdot\left(\td{\vec{x}}{t}j\right)
+ \pd{}{\mu}\left(\td{\mu}{t}j\right) + \pd{}{p}\left(\td{p}{t}j\right) \\
= \pd{}{\mu}\left(D_{\mu\mu}\pd{j}{\mu}\right) 
+ \nabla\cdot\left(\mathbf{D}_\perp \cdot\nabla j\right) + S.
\end{multline}

with 

\begin{align}
\td{\vec{x}}{t} =& 
\vec{V}_{\rm sw}+\mu v\vec{b}, \label{eq:fte_x}\\
\td{\mu}{t}=&
\frac{1-\mu^2}{2}\left(v \nabla\cdot\vec{b} + \mu \nabla\cdot\vec{V}_{\rm sw} - 3 \mu \vec{b}\vec{b}:\nabla\vec{V}_{\rm sw}- \frac{2}{v}\vec{b}\cdot\td{\vec{V}_{\rm sw}}{t} \right),\label{eq:fte_mu}\\
\td{p}{t} =&
 \left(\frac{1-3\mu^2}{2}(\vec{b}\vec{b}:\nabla\vec{V}_{\rm sw}) - \frac{1-\mu^2}{2}\nabla\cdot\vec{V}_{\rm sw}-\frac{\mu }{v}\vec{b}\cdot\td{\vec{V}_{\rm sw}}{t}\right) p. \label{eq:fte_p} 
\end{align}

In this equation, $j(\Vec{x}, p,\mu, t)$ represents the gyro-averaged differential particle intensity, which varies with time $t$, spatial coordinate $\vec{x}$, momentum magnitude $p$, and pitch-angle cosine $\mu$. Moreover, $j(E)$, is related to the gyro-averaged particle distribution $f$ through $j(E) = p^2f(p)$, with $E$ denoting the particle energy. The remaining terms in Equation~\eqref{eq:fte} include a source term $S$, the particle speed $v$, the solar wind velocity $\vec{V}_{\rm sw}$, the unit vector $\vec{b}$, indicating the direction of the ambient interplanetary magnetic field. In this work, $\vec{V}_{\rm sw}$ and $\Vec{b}$ are derived from the MAS and EUHFORIA models.

The two diffusion processes on the right-hand side of Equation~\eqref{eq:fte} represent pitch-angle diffusion, characterized by $D_{\mu\mu}$, and spatial diffusion perpendicular to the background magnetic field, described by the tensor $\mathbf{D}_\perp = D_\perp (I- \Vec{b}\Vec{b})$, where $I$ represents the identity matrix. These diffusion mechanisms account for the influence of fluctuating magnetic fields $\delta \vec{B}$ on particle transport. In this work, we assume the magnetic turbulence to be magnetostatic, incompressible, and decomposable into a slab and a 2D component, as described by e.g., \citet{Matthaeus1990JGR....9520673M}. Furthermore, similar to \citet{wijsen2023JGRA}, we assume that  the ratio of magnetic variance to the background magnetic field magnitude is expressed by the following relationship:

\begin{equation} \label{eq:varB_radialDependence}
  \frac{\delta B^2}{B_0^2} =
  \begin{cases}
  \Lambda_0 \left(\frac{r}{r_0}\right)^{\alpha_1} &  r  \leq r_1 = 0.5\text{ au}\\ 
  \Lambda_1  \left(\frac{r}{r_0}\right)^{\alpha_2} & r_1 < r  \leq r_2 = 2.0 \text{ au} \\ 
  \Lambda_2   & r_2 < r
  \end{cases}   
\end{equation}

Here, $B_0$ represents the average background interplanetary magnetic field, with $r_0 = 0.1$~au,  $\Lambda_0 = 0.1$, 
$\Lambda_1 = \Lambda_0 (r_1/r_0)^{\alpha_1-\alpha_2} \approx 0.15$, $\Lambda_2 = \Lambda_1 (r_2/r_0)^{\alpha_2} \approx 0.32$,   where $\alpha_1 = 0.5$, and $\alpha_2 = 0.25$. 

The choice of Eq.~\eqref{eq:varB_radialDependence} aims to capture the varying radial dependence of the background magnetic field and the turbulent field $\delta B$. Within $0.5$~au, the average background interplanetary magnetic field $\vec{B}_0$ is predominantly radial and scales on average as $\sim r^{-2}$; beyond $2$~au, $\vec{B}_0$ becomes predominantly azimuthal and scales on average as $\sim r^{-1}$. Recent observations from PSP regarding the turbulent $\delta B^2$ suggest a dependence of $\sim r^{-3.5}$ within $1$~au \citep{chhiber2022ApJ}. Beyond 1~au, the radial dependence exhibits a hardening and even reverses \citep{Adhikari_2023}. Furthermore, following \citet{bieber1994}, we assume that $\delta B^2_{\rm slab} = 0.2\delta B^2$ and $\delta B^2_{\rm 2D} = 0.8\delta B^2$ throughout the simulation domain. The correlation lengths for the 2D and slab components are prescribed as $\ell_{2D} = (0.0074~\text{au}) (r / 1~\text{au})^{1.1}$ and $\ell_{\text{slab}} = 3.9\times\ell_{2D}$, respectively \citep{strauss2017,weygand2011}.

The functional form of the pitch-angle diffusion coefficient is derived from quasi-linear theory and scaled such that the parallel mean free path equals \citep{teufel2003}:

\begin{equation}
\label{eq:LambdaParl}
\lambda_{\parallel} = \frac{3 s}{\pi (s - 1)} \ell_{\rm slab} R^2 \frac{B_0^2}{\delta B_{\rm sl}^2} \left[ \frac{1}{4} + \frac{2 R^{-s}}{(2 - s)(4 - s)} \right]
\end{equation}

Here, $s = 5/3$ represents the spectral index of a Kolmogorov turbulence power spectrum, and $R = R_L/\ell_{\rm slab}$, with $R_L$ being the Larmor radius for a $90^{\circ}$ pitch angle.  The parallel mean free path is related to the pitch angle diffusion coefficient through $D_{\mu\mu}$ through \citep{hasselmann70}

 \begin{equation}
\lambda_\parallel = \frac{3v}{8}\int_{-1}^{1}\frac{\left( 1 - \mu^2\right)^2}{D_{\mu\mu}}d\mu.
\end{equation}

The cross-field diffusion coefficient utilized in the PARADISE simulations is based on the nonlinear guiding center (NLGC) theory of \citet{matthaeus2003}, modified by \citet{engelbrecht2019} to explicitly incorporate a dependence on the particle's pitch angle:

\begin{equation}\label{eq:nlgc}
    D_\perp =\mu^2 v \lambda_\parallel^{1/3}\left( a^2\sqrt{3\pi}\frac{2\nu -1 }{\nu} \frac{\Gamma(\nu)}{\Gamma(\nu -1/2)} \frac{\delta B^2_{2D}}{B_0^2} \ell_{2D} \right)^{2/3}
\end{equation}

Here, $\nu = s/2$, and $\Gamma$ denotes the gamma function. The free parameter $a$ is set to $\sqrt{1/3}$, consistent with the findings of \citet{matthaeus2003}. 

In this work, we assume that the shock continuously emits energetic particles. Assuming the shock to be an MHD discontinuity, the source function $S$ can then be expressed as:

\begin{equation}
S(\vec{r}, p, t) = Q(p, t)\delta(\vec{r} - \vec{r}_\mathrm{s}),
\end{equation}

Here, $Q(p, t)$ represents the particle emission rate, and $\vec{r}_\mathrm{s}(\vec{r}, t)$ describes the evolution of the shock front in space and time. 
We make the assumption that the emission rate follows the functional form:

\begin{equation}\label{eq:emission_rate}
Q(p) = u_1 \frac{r_\mathrm{c} - 1}{r_\mathrm{g}} \frac{p^3}{3} \frac{\partial f_s}{\partial p},
\end{equation}

Here, $u_1$ denotes the upstream solar wind speed measured in the shock frame and $f_\mathrm{s}(p)$ represents the isotropic particle distribution function at the shock and is given by Eq.~\eqref{eq:DSA}. 

As noted before, the solar wind velocity \(\mathbf{v}_{\text{sw}}\), magnetic field \(\mathbf{B}\), and proton density $n_p$ are derived from the MHD models MAS and EUHFORIA. MAS provides the MHD background in the corona (up to 0.1 au), while EUHFORIA models the solar wind beyond 0.1 au. The outer boundary of EUHFORIA is set at 3 au, ensuring no interference with the SEP simulations performed using PARADISE. We employ the MHD variables calculated by MAS at 21.5 \(R_\odot\) as inner boundary conditions for EUHFORIA, enabling a continuous MHD description of the inner heliosphere, from the solar surface to 3 au.



\end{document}